\documentclass[12pt]{amsart}
\usepackage{fullpage,amsthm}
\usepackage{amsmath,amsfonts,amssymb,subfigure} 
\usepackage{graphicx}
\pdfoutput=1

\input xy
\xyoption{all}

\newtheorem{thm}{Theorem}
\newtheorem{corl}[thm]{Corollary}
\newtheorem{lma}[thm]{Lemma} 
\newtheorem{prop}[thm]{Proposition}
\newtheorem{defn}[thm]{Definition}

\newtheorem{ex}[thm]{Example}
\newtheorem{rem}[thm]{Remark}

\def\bA{\mathbb{A}}

\DeclareMathOperator{\Ad}{Ad}

\def\Aut{\mathrm{Aut}}

\def\BB{\mathbb{B}}
\def\bar{\overline}

\def\C{\mathbb{C}}
\def\cC{\mathcal{C}}

\def\cO{\mathcal{O}}

\def\G{\mathcal{G}}

\def\H{\mathcal{H}}
\def\bH{\mathbb{H}}

\DeclareMathOperator{\Hom}{Hom}
\DeclareMathOperator{\Hol}{Hol}

\def\id{\mathrm{id}}

\DeclareMathOperator{\Inv}{Inv}
\def\into{\hookrightarrow}

\def\nn{\nonumber}

\def\cP{\mathcal{P}}

\def\R{\mathbb{R}}

\def\S{\mathcal{S}}

\DeclareMathOperator{\tr}{Tr}
\def\tilde{\widetilde}

\def\U{\mathcal{U}}
\def\u{\mathfrak{u}}

\def\X{\mathcal{X}}

\def\Z{\mathbb{Z}}

\renewcommand{\cP}{{\mathcal P}}

\title{Gauge networks in noncommutative geometry}
\author{Matilde Marcolli}
\address{Division of Physics, Mathematics, and Astronomy, California Institute of Technology, 1200 E California Blvd, Pasadena, CA 91125, USA}
\author{Walter D. van Suijlekom}
\address{Institute for Mathematics, Astrophysics and Particle Physics,
Radboud University Nijmegen, Heyendaalseweg 135, 6525 AJ Nijmegen, The Netherlands}
\date{15 January 2013}

\begin{document}

\begin{abstract}
We introduce gauge networks as generalizations of spin networks and lattice gauge fields to almost-commutative manifolds. The configuration space of quiver representations (modulo equivalence) in the category of finite spectral triples is studied; gauge networks appear as an orthonormal basis in a corresponding Hilbert space. We give many examples of gauge networks, also beyond the well-known spin network examples. We find a Hamiltonian operator on this Hilbert space, inducing a time evolution on the $C^*$-algebra of gauge network correspondences. 

Given a representation in the category of spectral triples of a quiver embedded in a spin manifold, we define a discretized Dirac operator on the quiver. We compute the spectral action of this Dirac operator on a four-dimensional lattice, and find that it reduces to the Wilson action for lattice gauge theories and a Higgs field lattice system. As such, in the continuum limit it reduces to the Yang--Mills--Higgs system. For the three-dimensional case, we relate the spectral action functional to the Kogut--Susskind Hamiltonian.

\end{abstract}
\maketitle

\section{Introduction}

We develop a formalism of {\em gauge networks} that bridges between three
apparently different notions: the theory of spin networks in quantum gravity, 
lattice gauge theory, and the almost-commutative geometries used in the
construction of particle physics models via noncommutative geometry. 

\smallskip

The main idea behind the spin networks approach to quantum gravity is that
a space continuum is replaced by quanta of space carried by the vertices of 
a graph and quanta of areas, representing the boundary surface between
two adjacent quanta of volume, carried by the graph edges. The metric data
are encoded by holonomies described by $SU(2)$ representations associated
to the edges with intertwiners at the vertices, \cite{Bae94}, \cite{Bae98}.

\smallskip

On the other hand, in the noncommutative geometry approach to models
of matter coupled to gravity, one considers a non-commutative geometry that
is locally a product of an ordinary $4$-dimensional spacetime manifold and
a {\em finite spectral triple}. A spectral triple, in general, is a noncommutative
generalization of a compact spin manifold, defined by the data $(A,H,D)$ of
an involutive algebra $A$ with a representation as bounded operators
on a Hilbert space $H$, and a Dirac operator, which is a densely defined
self-adjoint operator with compact resolvent, satisfying the compatibility
condition that commutators with elements in the algebra are bounded. In
the finite case, both $A$ and $H$ are finite dimensional: such a space
corresponds to a metrically zero dimensional noncommutative space. 
A product space of a finite spectral triple and an ordinary manifold (also 
seen as a spectral triple) is known as an {\em almost-commutative geometry}. There is a natural action functional, the {\em spectral
action}, on such spaces, whose asymptotic expansion recovers the classical
action for gravity coupled to matter, where the matter sector Lagrangian is 
determined by the choice of the finite noncommutative space,  \cite{CC96}, \cite{CC10},
\cite{CCM07}, \cite{CM07}.

\smallskip

Just as the notion of a spin network encodes the idea of a discretization
of a 3-manifold, one can consider a similar approach in the case of the
almost-commutative geometries and ``discretize" the manifold part of
the geometry, transforming it into the data of a graph, with finite spectral
triples attached to the vertices and morphisms attached to the edges.
This is the basis for our definition of gauge networks, which can be
thought of as {\em quanta of noncommutative space}. While we mostly
restrict our attention to the gauge case, where the Dirac operators
in the finite spectral triples are trivial, the same construction works more
generally. We show that the manifold Dirac operator of the almost-commutative geometry can be replaced by a discretized version defined
in terms of the graph and of holonomies along the edges. 

\smallskip

In lattice gauge theory, the Wilson action defined in terms of holonomies
recovers, in the continuum limit, the Yang--Mills action, \cite{Cre83}.
We show that the
spectral action of the Dirac operator on a gauge network recovers the
Wilson action with additional terms that give the correct action for a lattice
gauge theory with a Higgs field in the adjoint representation, \cite{DJK84}, \cite{LRV81}.

\smallskip

In Section \ref{quiverSec} we construct a category whose objects are {\em finite
spectral triples} and whose morphisms are pairs of an algebra morphism and
a unitary operator with a compatibility condition, and a subcategory made of  
those finite spectral triples that have trivial Dirac operator. We give some explicit
examples, including those related to Yang--Mills theory and to the Standard Model. 
Using the Artin--Wedderburn theorem, one can write the algebras as sums of
matrix algebras and describe the morphisms in terms of Bratteli diagrams and
of more general braid Bratteli diagrams, which keep into account the permutations 
of blocks of the same dimension. We then introduce the main objects of our
constructions, which are representations of quivers (oriented graphs) in the category
of finite spectral triples described above. The configuration space ${\mathcal X}$ is
the space of such representations and we also consider its quotient by a natural
group ${\mathcal G}$ of symmetries given by the invertible morphisms at each vertex of the graph. This quotient can be understood as taking equivalence classes of quiver representations in the category of finite spectral triples. The space ${\mathcal X}$ and the ${\mathcal G}$-invariants of 
$L^2({\mathcal X})$ are described more explicitly using the orbit-stabilizer theorem, the Peter--Weyl theorem for compact Lie groups, and its extension to homogeneous spaces. An orthonormal basis is given in terms
of the intertwiners at vertices. Thus, the data of a {\em gauge network} can be
defined in terms of a quiver representation in the category of finite spectral triples 
with vanishing Dirac operator, carrying unitary Lie group representations along 
the edges and intertwiners at vertices.  We show that
the data obtained in this way, in the case where the pair $(A_v,H_v)$ at each vertex
is $(M_N(\C), \C^N)$ with trivial Dirac operator, recovers the case of 
$U(N)$ spin networks.  Other examples of gauge networks are discussed in this
section, including abelian spin networks, $U(N)$ spin networks, and some non-spin-network
examples with trivial Hilbert space (the representation in the spectral triple datum
is not assumed to be faithful), where the Peter--Weyl decomposition of 
$L^2({\mathcal X})$  can be described in terms of Gelfand--Tsetlin diagrams. 

\smallskip

In Section \ref{corrSec} we give a categorical formulation by introducing
morphisms between gauge networks in the form of correspondences defined
by bimodules. We also define a $C^*$-algebra of gauge network 
correspondences, and a time evolution, where the Hamiltonian is an operator
on $L^2({\mathcal X})$ defined as a sum of quadratic Casimir operators 
of the Lie groups ${\mathcal U}(A_{t(e)})$. This makes the noncommutative
geometries described by gauge networks dynamical.

\smallskip

In Section \ref{latticeSec} we introduce a notion of (discretized) Dirac operator for 
a representation (in the category of spectral triples) of a quiver embedded in a Riemannian spin manifold, and we show that in
the lattice case, in the continuum limit where the lattice size goes to zero, this
recovers the usual geometric Dirac operator on a manifold. We also consider
Dirac operators twisted by gauge potentials. These Dirac operators turn the
quiver representations into spectral triples. We then consider the spectral action, computed for a quiver that is a four-dimensional lattice.
We show that it reduces to the Wilson action for lattice gauge theory
and a Higgs field lattice system, with the Higgs field in the
adjoint representation. In the case of a 3-dimensional lattice we recover the 
Kogut--Susskind Hamiltonian. We finish the section with a proposal for
an extension of our formalism from gauge networks to gauge foams, 
which we hope to return to in future work.

\subsection*{Acknowledgements}
The first author is partially supported by NSF grants DMS-0901221, DMS-1007207, 
DMS-1201512, and PHY-1205440.
The second author is supported in part by the ESF Research Networking Programme ``Low-Dimensional Topology and Geometry with Mathematical Physics (ITGP)''.

\section{Quiver representations and finite spectral triples}\label{quiverSec}
We introduce the notion of a gauge network, thereby generalizing spin networks to quanta of {\it noncommutative} space.
We adopt a (noncommutative) differential geometrical point of view and take spectral triples as our starting point.

\subsection{Finite-dimensional algebra representations and finite spectral triples}
We start by introducing a category of finite-dimensional algebras, together with a representation on a Hilbert space. 

\begin{defn}
The category $\cC_0$ has as objects triples $(A,\lambda,H)$ where $A$ is a finite-dimensional (unital, complex) $*$-algebra, and $\lambda$ is a $*$-representation on an inner product space $H$.
A morphism in $\Hom((A_1,H_1),(A_2,H_2))$ is given by a pair $(\phi,L)$ consisting of a unital $*$-algebra map $\phi: A_1 \to A_2$ and a unitary $L: H_1\to H_2$ such that
\begin{equation}
\label{eq:equivariance}
L \lambda_1(a)L^* = \lambda_2(\phi(a)) ; \qquad (\forall a \in A_1).
\end{equation} 
\end{defn}



An alternative definition of the above category $\cC_0$ is as a category of finite spectral triples $(A,H,D)$ with vanishing Dirac operator $D=0$. 
\begin{defn}
The category $\cC$ has as objects finite spectral triples $(A,\lambda,H,D)$,\footnote{If no confusion can arise, we will also write $(A,H,D)$ for such a spectral triple.} {\it i.e.} $A$ is a finite-dimensional (complex) $*$-algebra, $H$ is an inner product space on which $A$ acts involutively via $\lambda$, and $D$ is a symmetric linear operator on $H$ (referred to as finite Dirac operator). 
A morphism in $\Hom((A_1,H_1,D_1),(A_2,H_2,D_2))$ is given by a 
pair $(\phi,L)$ consisting of a unital $*$-algebra map $\phi: A_1 \to A_2$ and an unitary $L: H_1\to H_2$ such that Eq. \eqref{eq:equivariance} holds, as well as
\begin{equation}
\label{eq:equivariance-D}
L D_1 L^* = D_2.
\end{equation}
\end{defn}

Note that in particular, $\cC_0 \subset \cC$ is a full subcategory. 
In contrast to Mesland's category \cite{Mesland} of spectral triples, here we only take {\it correspondences} that are induced by the algebra map $\phi:A_1 \to A_2$, whilst also explicitly including the compatible unitary map $L:H_1 \to H_2$. In fact, ${}_{\phi(A_1)} (A_2)_{A_2}$ is an $A_1-A_2$-bimodule where $A_1$ acts via the map $\phi:A_1 \to A_2$, and for which the above unitary implements
$$
H_1\simeq A_2 \otimes_{A_2} H_2 \simeq H_2,
$$
compatibly with the action of $A_1$. The difference $D_2-D_1$ can be captured by a connection on the bimodule ${}_{\phi(A_1)} (A_2)_{A_2}$.
For us, allowing for all correspondences between finite spectral triples yields a slightly too large category. However, it is an interesting question how that would generalize the gauge networks that are introduced below. 



Let us analyze the structure of the morphisms in the category $\cC_0$ (or, which is the same, in $\cC$ with vanishing $D$). We start with some illustrative examples.

\begin{ex}
\label{ex:hom0}
Suppose $A_1 = M_N(\C)=A_2$, $H_1=\C^N=H_2 $. A unital $*$-algebra map $\phi:A_1 \to A_2 $ is given by 
$$
m \in M_N(\C) \mapsto u m u^* \in M_N(\C). 
$$
for a unitary $N \times N$ matrix $u$. A compatible unitary map $L:H_1 \to H_2$ is given by the same unitary matrix $u$. We conclude that $\Hom((A_1,H_1),(A_2,H_2)) \simeq U(N)$. Later, we will see that this example lies at the basis of $U(N)$ Yang--Mills theory. 
\end{ex}

\begin{ex}
\label{ex:hom-SM}
Suppose $A_1 = \C \oplus M_2(\C) \oplus M_3(\C)=A_2$, $H_1 = \C \oplus \C^2 \oplus \C^3$. A unital $*$-algebra map $\phi:A_1 \to A_2 $ is given by conjugation with the unitary map
$$
(v_1, v_2, v_3) \in \C \oplus \C^2 \oplus \C^3 \mapsto 
(u_1 v_1, u_2 v_2, u_3 v_3) \in \C \oplus \C^2 \oplus \C^3
$$ 
where $(u_1,u_2,u_3) \in U(1) \times U(2) \times U(3)$. Thus, $\Hom((A_1,H_1),(A_2,H_2)) \simeq U(1) \times U(2) \times U(3)$. This example is closely related to the noncommutative description of the Standard Model \cite{CCM07}. 
\end{ex}

\begin{ex}
\label{ex:hom1}
Suppose $A_1 = \C$, $H_1=\C \oplus \C$, and $A_2 = \C \oplus M_2(\C)$, $H_2 =\C^2$ ({\it i.e.} $\C \subset A_2 $ acts trivially on $H_2$). The only unital $*$-algebra map $\phi:A_1 \to A_2 $ is given by 
$$
 z  \in \C \mapsto \begin{pmatrix}  z  &  \\ &  z  1_2 \end{pmatrix}
\in  \C \oplus M_2(\C). 
$$
The linear map $L$ is given by any unitary $2 \times 2$ matrix, for which automatically $L \lambda_1( z ) L^* = \lambda_2( z )$, or, explicitly
$$
L ( z  \oplus  z ) L^* =  z  1_2. 
$$ 
Thus, in this case $\Hom((A_1,H_1),(A_2,H_2)) \simeq U(2)$.
\end{ex}

\begin{ex}
\label{ex:hom2}
Suppose $A_1 = \C \oplus M_2(\C), H_1 = \C \oplus \C^2$, and $A_2 = M_3(\C), H_2 = \C^3$. A unital $*$-algebra map $\phi: A_1 \to A_2$ is then of one of the following two forms:
\begin{align}
( z , a) \in \C \oplus M_2(\C) &\mapsto u \begin{pmatrix}  z  &  \\ & a \end{pmatrix} u^* \in M_3(\C)
\tag{a}
\intertext{where $u \in U(3)$, or, with kernel $M_2(\C)$:}
( z , a) \in \C \oplus M_2(\C) &\mapsto  z  1_3 \in M_3(\C)
\tag{b}
\end{align}
In both cases, a unitary map from $H_1$ to $H_2$ is given by
$$
(x,y) \in \C \oplus \C^2 \mapsto U \begin{pmatrix} x \\ y \end{pmatrix}  \in \C^3
$$
with $U \in U(3)$. Let us first consider the case \emph{(a)}; then Eq. \eqref{eq:equivariance} demands
$$
u \begin{pmatrix}  z  &  \\ & a \end{pmatrix} u^* = 
U \begin{pmatrix}  z  &  \\ & a \end{pmatrix} U^*.
$$
so that the $*$-algebra map $\phi$ can equally well be implemented using the unitary $U$ instead of $u$.

For the case \emph{(b)} the equivariance demands
$$
 z  1_3 = U \begin{pmatrix}  z  &  \\ & a \end{pmatrix} U^*.
$$
which cannot be satisfied for arbitrary $(z,a) \in A_1$. 

Thus, we have in this case $ \Hom((A_1,H_1),(A_2,H_2))\simeq U(3)$.
\end{ex}

More generally, by the Artin--Wedderburn Theorem any finite-dimensional $*$-algebra is a direct sum of matrix algebras:
\begin{equation}
\label{eq:wedderburn}
A_1 \simeq \bigoplus_{i=1}^k M_{N_i} (\C); \qquad A_2 \simeq \bigoplus_{j=1}^{k'} M_{N'_j} (\C),
\end{equation}
for some (not necessarily different) integers $N_1,\ldots,N_k$. 

Upon fixing the above isomorphisms, any unital $*$-algebra map $\phi:A_1 \to A_2$ can be written as the direct sum of representations:
$$
\phi_j :  \bigoplus_{i=1}^k M_{N_i} (\C) \to M_{N'_j} (\C).
$$ 
Moreover, $\phi_j$ splits as a direct sum of representation $\phi_{ij} : M_{N_i} (\C) \to M_{N'_j} (\C)$ with multiplicity $d_{ij} \geq 0$. These multiplicities fulfill
$$
N_j' = \sum_i d_{ij} N_i.
$$

\begin{figure}
\includegraphics{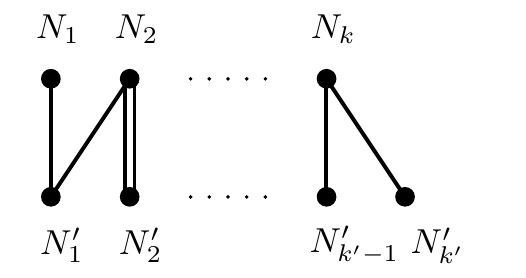}
\caption{Bratteli diagram $\BB$ for a unital $*$-algebra map $A_1 \to A_2$.}
\label{fig:Bratteli}
\end{figure}

\begin{figure}
\includegraphics{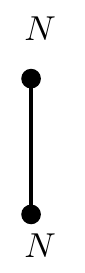}
\caption{The only Bratteli diagram $\BB$ for unital $*$-algebra maps $\phi: M_N(\C) \to M_N(\C)$ and the isometries $L: \C^N \to \C^N$ (Example \ref{ex:hom0}).}
\label{fig:Bratteli-ex0}
\end{figure}

\begin{figure}
\includegraphics{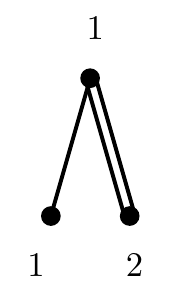}
\caption{The Bratteli diagram for the $*$-algebra maps $\C  \to \C \oplus M_2(\C)$ of Example \ref{ex:hom1}.}
\label{fig:Bratteli-ex1}
\end{figure}

\begin{figure}
\subfigure[]{\includegraphics{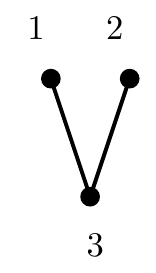}}
\subfigure[]{\includegraphics{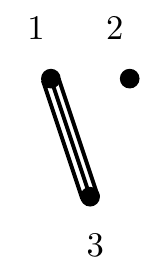}}
\caption{Two Bratteli diagrams for the $*$-algebra maps $\C \oplus M_2(\C) \to M_3(\C)$ of Example \ref{ex:hom2}.}
\label{fig:Bratteli-ex2}
\end{figure}

\begin{figure}
\includegraphics{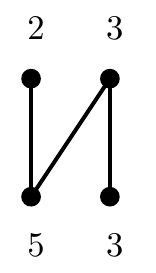}
\caption{Bratteli diagram for the only unital $*$-algebra map $M_2 (\C) \oplus M_3(\C) \to M_5(\C) \oplus M_3(\C)$ given $(a,b) \mapsto (a\oplus b, b)$.}
\label{fig:Bratteli-ex3}
\end{figure}

This can be nicely depicted in a so-called {\it Bratteli diagram} $\BB$ for the pair $(A_1,A_2)$ \cite{Bra72}. It consists of two rows of vertices, the top row consisting of $k$ vertices, labeled by $N_1, \ldots ,N_k$, and the bottom row consisting of $k'$ vertices, labeled by $N_1', \ldots ,N_{k'}'$ (cf. Figure \ref{fig:Bratteli}). Then, between vertex $i$ (top row) and $j$ (bottom row) there are precisely $d_{ij}$ edges. Since the $*$-algebra map $A_1 \to A_2$ is unital, all vertices in the bottom row are reached by an edge, but the top row might have vacant vertices (cf. Figure \ref{fig:Bratteli-ex0}, \ref{fig:Bratteli-ex1}, \ref{fig:Bratteli-ex2} and \ref{fig:Bratteli-ex3} for examples of Bratteli diagrams).

Conversely, any such diagram $\BB$ (for the pair $(A_1,A_2)$) gives rise to a morphism $\phi_\BB: A_1 \to A_2$ by simply embedding the matrix blocks of $A_1$ into those of $A_2$, following the lines in $\BB$. All other unital $*$-algebra morphisms $\phi: A_1 \to A_2$ can be obtained from $\phi_\BB$ after a change of basis: $\phi(\cdot) = U \phi_\BB(\cdot) U^* =: \Ad U \phi_\BB(\cdot)$ for some unitary $U$ in $A_2$.


\begin{lma}
\label{lma:splitting-phi}
Let $(\phi,L)$ be a morphism in $\Hom((A_1,H_1),(A_2,H_2))$ and write $A_1 = \tilde A_1 \oplus \ker \lambda_1$ and $A_2 = \tilde A_2 \oplus \ker \lambda_2$. Then $\phi= \tilde \phi + \phi_0$ where $\tilde \phi: \tilde A_1 \to \tilde A_2$ and $\phi_0 : A_1 \to \ker \lambda_2$ are $*$-algebra maps such that
$$
\tilde \phi(\tilde a) = L \tilde a L^* \qquad ( \tilde a \in \tilde A_1),
$$
where we have identified $\tilde A_i \simeq \lambda_i(A_i)$.
\end{lma}
\proof
Since $\ker \lambda_i$ ($i=1,2$) is a two-sided $*$-ideal in $A_i$, it is a direct sum of some of the matrix algebras in the decomposition \eqref{eq:wedderburn} of $A_i$; the complement $\tilde A_i$ is the direct sum of the remaining matrix algebras. Thus, we can write according to this decomposition:
$$
\phi(\tilde a, a_0) = \tilde \phi(\tilde a, a_0) + \phi_0(\tilde a, a_0).
$$ 
The equivariance condition \eqref{eq:equivariance} reads
$$
\lambda_2( \tilde \phi(\tilde a,a_0)) = L \lambda_1(\tilde a) L^*
$$
so that $\tilde \phi(\tilde a,a_0)\equiv \tilde \phi(\tilde a)$, independent of $a_0$.  
\endproof

Note that the map $\tilde \phi$ is thus necessarily injective (though $\phi_0$ need not be so); this explains why Example \ref{ex:hom2}(b) was not allowed. 

The map $\phi$ can thus be described by two subdiagrams in the Bratteli diagrams $\BB$ for $\phi$: a subdiagram (called $\tilde \BB$) for the $*$-algebra map $\tilde \phi: \tilde A_1 \to \tilde A_2$ and one (called $\BB_0$) for the  $*$-algebra map $\phi_0: A_1 \to \ker A_2$. More precisely, the integers $N_i$ and $N_j'$ corresponding to $\tilde A_1$ and $\tilde A_2$, respectively, appear at the top and bottom row vertices in $\tilde \BB$, while the integers corresponding to $A_1$ and $\ker \lambda_2$ label the respective top and bottom row of vertices in the diagram $\BB_0$. Injectivity of $\tilde \phi$ implies that the Bratteli diagram $\tilde \BB$ has no vacant vertices in the top row. 
\begin{ex}
Consider the algebra map of Example \ref{ex:hom1}:
$$
\phi( z ) =  z  \oplus  z  1_2 =: \phi_0( z ) \oplus \tilde \phi( z ),
$$
with corresponding Bratteli diagram $\BB$ in Figure \ref{fig:Bratteli-ex1}. The Bratteli subdiagrams $\BB_0$ and $\tilde \BB$ are given in Figure \ref{fig:subBratteli-ex1}.

\begin{figure}
\subfigure[$\BB$]{\includegraphics{Bratteli112.pdf}}
\subfigure[$\tilde \BB$]{\includegraphics{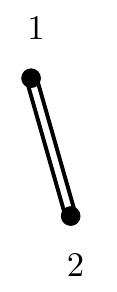}}
\subfigure[$\BB_0$]{\includegraphics{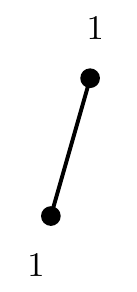}}
\caption{The Bratteli diagram $\BB$ for the $*$-algebra map $\phi: \C  \to \C \oplus M_2(\C)$ of Example \ref{ex:hom1} and the corresponding subdiagrams $\tilde \BB$ and $\BB_0$, corresponding to the respective algebra maps $\tilde \phi : \C \to M_2(\C)$ and $\phi_0 : \C \to \C$ that satisfy $\phi=\tilde \phi + \phi_0$.}
\label{fig:subBratteli-ex1}
\end{figure}
\end{ex}

We introduce the following unitary subgroup of the group of unitaries of $H$, with respect to a faithful representation of $\tilde A$ on $H$:
\begin{equation}
\Aut_{\tilde A}(H) := \{ U \in \U(H): U \tilde A U^* = \tilde A \}. 
\end{equation}
Note that such groups have been considered also in \cite{LS01}. One can check that for the multiplicity $n$ representation of $M_N(\C)$ we have $\Aut_{M_N(\C)} (n\C^N) \simeq U(N) \times U(n)$. 

\begin{prop}
\label{prop:hom}
Let $A_1$ and $A_2$ be matrix algebras as above, with respective representation spaces $H_1$ and $H_2$
. In other words,
\begin{gather*}
A_1 = \bigoplus_{i=1}^{k+l} M_{N_i} (\C) ; \qquad A_2 = \bigoplus_{j=1}^{k'+l'} M_{N'_j} (\C),\\
H_1 = \bigoplus_{i=1}^k n_i \C^{N_i} ; \qquad H_2 = \bigoplus_{j=1}^{k'}  n_j' \C^{N_j'}
\end{gather*} 
Then any morphism $(\phi,L) \in \Hom( (A_1,H_1),(A_2,H_2))$ can be written as
\begin{equation}
\label{eq:action-isotropy}
\phi = \Ad U \phi_{\tilde \BB} + \Ad V \phi_{\BB_0} ; \qquad L = U L_{\tilde \BB}
\end{equation}
in terms of unitaries $U \in \Aut_{\tilde A_2}(H_2) 
$ and $V \in \U(\ker \lambda_2) \simeq \prod_{j=k'+1}^{k'+l'} U(N_j)$ and Bratteli diagrams $\tilde \BB, \BB_0$ for $*$-algebra maps $\tilde A_1 \into \tilde A_2$ and $A_1 \to \ker \lambda_2$, respectively. The unitary map $L_{\tilde \BB}: H_1 \to H_2$ implements the $*$-algebra map $\phi_{\tilde \BB} : \tilde A_1 \to \tilde A_2 $:
$$
L_{\tilde \BB} \tilde a L_{\tilde \BB}^* = \phi_{\tilde \BB}(\tilde a); \qquad (\tilde a \in \tilde A_1). 
$$
\end{prop}
\proof
The map $\tilde \phi: \tilde A_1 \to \tilde A_2$ induces a representation of $\tilde A_1$ on $H_2$. As before, there is a Bratteli diagram $\BB$ that dictates how the matrix blocks in $A_1$ embed in those of $A_2$. This means that an irreducible representation $\C^{N_j'}$ of $\tilde A_2$ decomposes as a direct sum of representation of $\tilde A_1$ according to the Bratteli subdiagram $\tilde \BB$ (with $d_{ij}$ lines between top vertex $i$ and bottom vertex $j$):
$$
\C^{N_j'} \simeq \bigoplus_i d_{ij} \C^{N_i},
$$
or, simply, $N_j' = \sum_i d_{ij} N_i$. This implies that
$$
\bigoplus_j n_j ' \C^{N_j'} \simeq \bigoplus_{i,j} n_j' d_{ij} \C^{N_i}.
$$
Now, the map $L: H_1 \to H_2$ is compatible with the representation of $A_1$ on both Hilbert spaces, so it maps each $n_i \C^{N_i}$ isometrically to $\bigoplus_{j} n_j' d_{ij} \C^{N_i}$. Thus, $n_i = \sum_j n_j' d_{ij}$.

Using this, we let $L_{\tilde \BB}$ be the unitary map that maps the standard bases of $n_i \C^{N_i}$ to that of $n_j'  d_{ij} \C^{N_i}$ inside $n_j' \C^{N_j}$. Any other such unitary map $L$ is then given after a change of basis in each $\C^{N_j'}$ by $L= U L_\BB$ with $U \in \Aut_{\tilde A_2}(H_2)$. 
By Lemma \ref{lma:splitting-phi}, the map $\tilde \phi: \tilde A_1 \to \tilde A_2$ is given by $\tilde\phi(\tilde a) = L \tilde a L^*$ so that also 
$$
\tilde \phi(\tilde a) = \Ad U \phi_{\tilde \BB} .
$$
in terms of $\phi_{\tilde\BB}(\tilde a) := L_{\tilde \BB} \tilde a L_{\tilde \BB}$. 
The remaining algebra map $\phi_0: A_1 \to \ker \lambda_2$ is given by a Bratteli diagram $\BB_0$ and a unitary $V \in \U(\ker \lambda_2)$ as $\phi_0 = \Ad V \phi_{\BB_0}$.
\endproof

\begin{figure}
\includegraphics{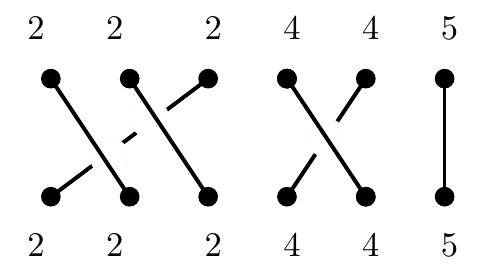}
\caption{An example of a `braid' Bratteli diagram, depicting permutations of matrix blocks of the same dimension in $M_2(\C)^{\oplus 3} \oplus M_4(\C)^{\oplus 2} \oplus M_5(\C)$.}
\label{fig:braid}
\end{figure}

\begin{corl}
\label{corl:iso}
Let $A$ be a matrix algebra, represented on $H$ as above. Then any isomorphism $(\alpha,U) \in \Aut( (A,H) )$ can be written as 
$$
\alpha = \Ad U \sigma_{\tilde \BB} + \Ad V \sigma_{\BB_0}; \qquad L = U \sigma_{\tilde \BB} 
$$
where $\sigma_{\tilde \BB}$ and $\sigma_{\BB_0}$ are products of permutations of matrix blocks in $A$ and correspondingly in $H$ of the same dimension (depicted by a `braid' Bratteli diagram $\BB$, as in Figure \ref{fig:braid}), and $U \in \Aut_{\tilde A}(H)$ and $V \in P\U(\ker \lambda)$, the projective unitary group.
\end{corl}
\proof
In order for $\phi$ to be surjective, both components $\tilde \phi: \tilde A \to \tilde A$ and $\phi_0 : A \to \ker \lambda$ should be surjective. Now, $\tilde \phi$ is already injective so that $\tilde\phi: \tilde A \to \tilde A$ is a $*$-automorphism. Since also $\tilde \phi(\tilde a) = L \tilde a L^*$, it follows necessarily that the unitary $L$ maps each $n_i \C^{N_i} \subset H$ to a $n_j\C^{N_j} \subset H$ with $N_i = N_j$ and $n_i = n_j$. This means that it is given by a permutation of blocks of the same dimension (depicted in a `braid' Bratteli diagram $\tilde \BB$) together with a unitary of these same dimensions, $U \in \Aut_{\tilde A} (H)$. But then also $\phi_0$ is a $*$-automorphism of $\ker \lambda$, hence given by a `braid' Bratteli diagram $\BB_0$, corresponding $\phi_{\BB_0}$ and a unitary $V \in \U(A)$. The reduction to the projective unitary group follows because the adjoint action of the center of $\U(\ker\lambda)$ on $\ker \lambda$ is trivial.
\endproof

We can thus identify $\Aut((A,\lambda,H)) \simeq \U(\tilde A) \rtimes S(\tilde A;H) \times  P\U(\ker \lambda) \rtimes S(\ker \lambda)$ where $S(\tilde A;H)$ denotes the group of permutations of the matrix blocks of equal dimension in $\tilde A$ and $H$, and similarly for $\ker \lambda$.


\subsection{Quiver representations}
Let $\Gamma$ be a directed graph. We describe gauge theories by considering $\Gamma$ as a {\it quiver} and represent it in the category $\cC$. Later, we will embed $\Gamma$ in a Riemannian spin manifold $M$, which is the `background' on which the gauge theory will be defined. 

\begin{defn}
A representation $\pi$ of a quiver $\Gamma$ in a category is an association of objects $\pi_v$ in that category to each vertex $v$ and morphisms $\pi_e$ in $\Hom(\pi_{s(e)}, \pi_{t(e)})$ to each directed edge $e$. 

Two representations $\pi, \pi' $ of $\Gamma$ in the same category are called equivalent if $\pi_v=\pi'_v$ for all $v \in \Gamma^{(0)}$ and if there exists a family of invertible morphisms $\phi_v \in \Hom( \pi(v),\pi(v))$ indexed by the vertices $v$ such that
$$
\pi_e = \phi_{t(e)} \circ \pi_e' \circ \phi_{s(e)}^{-1}
$$
\end{defn}
In other words, if we view a quiver $\Gamma$ as a category, a representation is simply given by a functor $\pi$ from $\Gamma$ to a category, and equivalent representation coincide on objects and are related via natural transformations.

In the case of the category $\cC$ (or $\cC_0$) a representation $\pi$ of the quiver $\Gamma$ assigns spectral triples $(A_v,H_v,D_v)$ (with $D_v = 0$ for $\cC_0$) to each vertex $v \in \Gamma^{(0)}$ and pairs $(\phi,L) \in \Hom( (A_{s(e)}, H_{s(e)}, D_{s(e)}), (A_{t(e)}, H_{t(e)}, D_{t(e)}))$ to each edge $e \in \Gamma^{(1)}$. 
We denote by $\X$ the space of such quiver representations $\pi : \Gamma \to \cC$. 
The collection of invertible morphisms $(\phi_v,L_v)$ for each vertex forms a group, which we denote by $\G$. 

We will now explicitly determine the form of the space $\X$ and the quotient $\X/\G$. For simplicity, we restrict to quiver representations in $\cC_0$ so that $D_v = 0$ for all vertices. 
 Recall from the previous section the decomposition $A_v = \tilde A_v \oplus \ker \lambda_v$ for each vertex $v$.
\begin{prop}
The space $\X$ of representations of $\Gamma$ in $\cC_0^s$ is 
\begin{equation*}
\X = \coprod_{\{A_v,H_v\}_v} \prod_{e \in \Gamma^{(1)}}\X_e
\end{equation*}
where $\{ A_v,H_v \}_v$ stands for the association $v \mapsto (A_v,\lambda_v, H_v)$ of an object in $\cC_0^s$ to each vertex $v \in \Gamma^{(0)}$, and 
$$
\X_e \simeq \coprod_{\BB_e} 
\Aut_{\tilde A_{t(e)}}(H_{t(e)}) \times 
\U( \ker \lambda_{t(e)} ) / \U( \ker \lambda_{t(e)} )_{\BB_{e0}}
$$
where $\BB_e$ is a Bratteli diagram with subdiagrams $\tilde \BB_e$ and $\BB_{e0}$ for each edge $e$, and $\U(\ker \lambda_{t(e)} )_{\BB_{e0}}$ is the isotropy subgroup in $\U(\ker \lambda_{t(e)} )$ of $\phi_{\BB_{e0}}$, acting according to \eqref{eq:action-isotropy}.
\end{prop}

\proof 
By definition $\X_e = \Hom( (A_{s(e)}, \lambda_{s(e)}, H_{s(e)}), (A_{t(e)}, \lambda_{t(e)},H_{t(e)}))$, after assigning an object $(A_v,\lambda_v,H_v)$ in $\cC_0$ to each vertex $v$. Proposition \ref{prop:hom} then shows that any element $(\phi_e,L_e) \in \X_e$ is of the form 
\begin{equation}
\label{eq:action-isotropy2}
\phi_e = \Ad U \phi_{\tilde \BB_e} + \Ad V \phi_{\BB_{e0}} ; \qquad L_e = U L_{\tilde \BB_e}
\end{equation}
in terms of unitaries $U \in \Aut_{\tilde A_{t(e)}}(H_{t(e)}),V \in \U(\ker \lambda_{t(e)})$ and a Bratteli diagram $\BB_e$ (with subdiagrams $\tilde \BB_e,\BB_{e0}$) for each edge $e$. In other words, the unitary group $\Aut_{\tilde A_{t(e)}}(H_{t(e)})$ together with the union of the $\U(\ker \lambda_{t(e)})$-orbits of $\phi_{\BB_{e0}}$ (acting as in \eqref{eq:action-isotropy2}) for all such $\BB_{e0}$ gives all of $\X_e$. Moreover, these orbits are disjoint because any pair $(\phi_e,L_e)$ uniquely determines a Bratteli diagram $\BB_e$ with subdiagrams $(\tilde \BB_e,\BB_{e0})$ for which $\phi_{e0} = \Ad V \phi_{\BB_{e0}}$. Thus, an application of the Orbit-stabilizer Theorem yields
$$
\X_e = \coprod_{\BB_e} \Aut_{\tilde A_{t(e)}}(H_{t(e)}) \times 
\U( \ker \lambda_{t(e)} ) / \U( \ker \lambda_{t(e)} )_{\BB_{e0}}
$$
for the isotropy subgroup  $\U( \ker \lambda_{t(e)})_{\BB_{e0}}$ of $\phi_{\BB_{e0}}$.
\endproof

We will denote an element in $\X$ by $(U_e,[V_e],\BB_e)_e$ where $U_e \in \Aut_{\tilde A_{t(e)}}(H_{t(e)})$ and $V_e \in \U(\ker \lambda_{t(e)})$. 

\begin{prop}
\label{prop:gauge-group}
Equivalences of quiver representations are determined by a collection
of unitaries $(g_v,\sigma_v) := (\tilde g_v,\tilde \sigma_v; g_{v0},\sigma_{v0}) \in \G_v := \Aut_{\tilde A_v}(H_v) \rtimes S(\tilde A_v;H_v) \times P\U(\ker \lambda_v) \rtimes S(\ker \lambda_v)$:
\begin{equation*}\label{GrepsUS}
\G \simeq \coprod_{\{A_v,H_v\}} \prod_{v \in \Gamma^{(0)}} \G_v,
\end{equation*}
where the action of $\{ (g_v,\sigma_v)_v\} \in \G$ on an element $(U_e,[V_e],\BB_e)_e   \in \X$ is given by
\begin{equation*}
\label{eq:gauge-transf}
(U_e, [V_e], \BB_e) \in \X_e \mapsto  (\tilde g_{t(e)} U_e \phi_{\tilde \BB_e}( \tilde g_{s(e)}^* ), [g_{t(e)0} V_e \phi_{\BB_{e0}}(g_{s(e)}^*) ], \sigma_{t(e)} \circ \BB_e \circ \sigma_{s(e)})
\end{equation*}
\end{prop}

\proof 
It follows from Corollary \ref{corl:iso} that automorphisms $(\phi_v,L_v) \in \Aut((A_v,H_v))$ are of the above form $(g_v, \sigma_v)$, with $\tilde \sigma_v$ given by the permutation of matrix blocks in $\tilde A_v$ of the same dimension, and similarly for $\ker \lambda_v$.
By definition, $(\phi_v,L_v) \in \G$ acts on $(\phi_e,L_e)_e \in \X$ as
$$
(\phi_e ,L_e) \mapsto (\phi_{t(e)} \phi_e \phi_{s(e)}^{-1} , L_{t(e)} L_e L_{s(e)}^{-1})
$$
Since the permutation $\sigma_{s(e)}$ and $\sigma_{t(e)}$ only act by interchanging matrix blocks of the same dimension, we will for simplicity restrict to the case $\sigma_v = \id$. Then, with $\phi_e = \Ad U_e \phi_{\tilde \BB_e}+ \Ad V_e \phi_{\BB_{e0}}$ and $L_e= U_e L_{\tilde \BB_e}$ we compute
\begin{align*}
\phi_{t(e)} \phi_e \phi_{s(e)}^{-1} &= 
\Ad \tilde g_{t(e)} \Ad U_e \phi_{\tilde \BB_e} \Ad g_{s(e)}^* 
+ \Ad g_{t(e)0} \Ad V_e \phi_{\BB_{e0}} \Ad g_{s(e)}^* \\
&=
\Ad \left( \tilde g_{t(e)} U_e \phi_{\tilde \BB_e} ( \tilde g_{s(e)}^* )\right ) \phi_{\tilde \BB_e}
+ \Ad \left( g_{t(e)0} V_e \phi_{\BB_{e0}}(g_{s(e)}^* )\right ) \phi_{\BB_{e0}}
\intertext{using the multiplicative property of the algebra maps $\phi_{\tilde \BB_e}$ and $\phi_{\BB_{e0}}$, respectively and the fact that $\phi_{\tilde \BB_e}(\tilde g_{s(e)}^*,g_{s(e)0}^*)$ only depends on $\tilde g_{s(e)}^*$. This agrees with}
L_{t(e)} L_e L_{s(e)}^{-1} &= \tilde g_{t(e)} U_e L_{\tilde \BB_e} \tilde g_{s(e)}^* = \tilde g_{t(e)} U_e \phi_{\tilde \BB_e}(\tilde g_{s(e)}^* )L_{\tilde \BB_e}.
\end{align*}
Thus, we obtain maps $U_e \mapsto \tilde g_{t(e)} U_e \phi_{\tilde \BB_e}(\tilde g_{s(e)}^*)$ and $V_e \mapsto g_{t(e)0} V_e \phi_{\BB_{e0}}(g_{s(e)}^*)$. The latter map is independent of the representative: if $\Ad V_e' \phi_{\BB_{e0}} = \phi_{\BB_{e0}}$ then since $g_{s(e)} \in A$ we have
$$
V_e V_e' \mapsto g_{t(e)0} V_e V_e' \phi_{\BB_{e0}}(g_{s(e)}^*) = g_{t(e)0} V_e \phi_{\BB_{e0}}(g_{s(e)}^*) V_e'.
$$
\endproof

\begin{ex}
The case where the data $(A,H)$ are given by $(A_v,H_v)=( M_N(\C),\C^N)$ reproduces the setting of $U(N)$ spin networks of \cite{Bae94} (see also Section \ref{sect:spin-networks} below). Indeed, the relevant (and only) Bratteli diagram is given in Figure \ref{fig:Bratteli-ex0}. The isotropy subgroup is trivial so that a quiver representation is an assignment of a unitary $u_e \in U(N)$ to each edge $e \in \Gamma^{(1)}$. The gauge group is given by an assignment of elements $g_v \in U(N)$ to each vertex $v \in \Gamma^{(0)}$ with the corresponding action given by
$$
u_e \to g_{t(e)} u_e g_{s(e)}^*. 
$$
\end{ex}

\subsection{Gauge networks}
The starting point for constructing a quantum theory is to construct a Hilbert space; inspired by \cite{Bae94}. It should be based on the classical configuration space, which in our case is $\X/\G$. As a union of homogeneous spaces for compact Lie groups, this is a measure space, equipped with products and sums of the Haar measures on the unitary groups. 
Hence, it makes sense to consider $L^2(\X)$. This Hilbert spaces carries an action of $\G$, induced by the action of $\G$ on $\X$. We aim for an explicit description of the space $L^2(\X/\G) \simeq L^2(\X)^\G$ of $\G$-invariant functions on $\X$. 

First, recall the Peter--Weyl Theorem for compact Lie groups, and its implication for homogeneous spaces.

\begin{thm}
\label{thm:PW}
Let $G$ be a compact Lie group. 
We have the following isomorphism of $G \times G$-representations:
$$
L^2(G) \simeq \bigoplus_{\rho \in \widehat{G}} \rho \otimes \rho^*
$$
with an element $(g_1,g_2) \in G \times G$ acting as
\begin{align*}
((g_1,g_2)f)(x) &= f(g_1^{-1} x g_2) ; \qquad (f \in L^2(G)),\\
(g_1,g_2) (y_1 \otimes y_2) &= \rho(g_1)y_1 \otimes \rho^*(g_2)(y_2); \qquad (y_1 \in \rho, y_2 \in \rho^*). 
\end{align*}
\end{thm}

\begin{corl}
\label{corl:PW}
Let $G$ be a compact Lie group, and $K$ and $H$ two mutually commuting closed subgroups. Then we have the following isomorphism of $G \times H$-representations:
$$
L^2(G/K) \simeq \bigoplus_{\rho \in \widehat{G}} \rho \otimes (\rho^*)^K
$$
where $\rho^K$ is the $K$-invariant subspace of the $G$-representation $\rho$. 
\end{corl}
\proof
We have $L^2(G/K) \simeq L^2(G)^K$, where $K$ acts via the embedding $\{ e \} \times K \subset G \times G$. For this action, we also have $(\rho \otimes \rho^*)^{\{e \} \times K} \simeq \rho \otimes (\rho^*)^K$. 
This is an isomorphism of $G \times H$-representations because the actions of $ \{e \} \times K$ and $G \times H$ commute.
\endproof

We apply this to our setting, where
\begin{equation}
\label{eq:L2-A}
L^2(\X) \simeq \bigoplus_{\{A_v,H_v\}}\bigotimes_e \bigoplus_{\BB_e} L^2\left(
\Aut_{\tilde A_{t(e)}}(H_{t(e)}) \times \U( \ker \lambda_{t(e)} ) / \U( \ker \lambda_{t(e)} )_{\BB_{e0}}\right)
\end{equation}
and with $\G$ acting according to Proposition \ref{prop:gauge-group}. We further condense notation by defining
$$
G_e := \Aut_{\tilde A_{t(e)}}(H_{t(e)}) \times \U( \ker \lambda_{t(e)} ); \qquad K_{\BB_e} := \{ e \} \times  \U( \ker \lambda_{t(e)} )_{\BB_{e0}},
$$
so that $L^2(\X) \simeq \bigoplus_{\{A_v,H_v\}}\bigotimes_e \bigoplus_{\BB_e}  L^2(G_e / K_{\BB_e})$.

\begin{prop}
There is an isomorphism of Hilbert spaces
$$
L^2(\X)  \simeq \bigoplus_{\{A_v,H_v\}}\bigotimes_e \bigoplus_{\BB_e} \bigoplus_{\rho_e \in \widehat{G_e}} \rho_e\otimes (\rho_e^*)^{K_{\BB_e}} 
$$
The group $\G$ acts accordingly on $L^2(\X)$:
$$ 
\bigoplus_{\{A_v,H_v\}}\bigotimes_e \bigoplus_{\BB_e} \bigoplus_{\rho_e \in \widehat{G_e}} 
 \rho_e(g_{t(e)}) \otimes \rho^*_e\circ\phi_\BB (g_{s(e)}),
$$
\end{prop}
\proof
This is an application of the above Corollary at each edge $e \in \Gamma^{(1)}$, with $G=G_e$, $K=K_{\BB_e}$ and $H = \phi_{\BB_e}(\U(A_{s(e)}))$. Indeed, the latter two groups mutually commute in $G$ since
$$
(e ,u') (\tilde \phi_\BB(u), \phi_{\BB_0}(u)) = (\tilde \phi_\BB(u), \phi_{\BB_0}(u)) (e,u') \iff \Ad u' \phi_{\BB_0}(u) = \phi_{\BB_0}(u); 
$$
for all $u \in \U( A_{s(e)}), u' \in K_{\BB_e} \equiv \U( \ker \lambda_{t(e)} )_{\BB_{e0}}$. This is true by the very definition of the stabilizer group $K_{\BB_e}$.
\endproof
Equivalently, we could associate first to each edge pairs $(\rho_{e}, \BB_e)$ of the above form, so that 
$$
L^2(\X) \simeq  \bigoplus_{\begin{smallmatrix}\{A_v,H_v\} \\ \{\rho_{e}, \BB_e\}  \end{smallmatrix}} \bigotimes_e  \rho_e\otimes (\rho_e^*)^{K_{\BB_e}} 
$$
and $\G$ acts accordingly:
$$
\bigoplus_{\begin{smallmatrix}\{A_v,H_v\} \\ \{\rho_{e}, \BB_e\}  \end{smallmatrix}} \bigotimes_e   \rho_e(g_{t(e)}) \otimes \rho^*_e\circ\phi_\BB (g_{s(e)})
$$
This leads to the following description of $L^2(\X)$:

\begin{prop}
\label{prop:gauge-network}
There is an isomorphism of Hilbert spaces
$$
L^2(\X) \simeq \bigoplus_{\begin{smallmatrix}\{A_v,H_v\} \\ \{\rho_{e}, \BB_e\}  \end{smallmatrix}} \bigotimes_v  \left(  \bigotimes_{e \in T(v)} \rho_{e} \otimes 
 \bigotimes_{e \in S(v)}  (\rho_{e}^*)^{K_{\BB_e}} \right),
$$
where $S(v)$ ($T(v)$) is the set of edges having $v$ as a source (target). The group $\G$ acts accordingly on $L^2(\X)$:
$$
\bigoplus_{\begin{smallmatrix}\{A_v,H_v\} \\ \{\rho_{e}, \BB_e\}  \end{smallmatrix}} \bigotimes_v 
\left(  \bigotimes_{e \in T(v)} \rho_{e} (g_v) \otimes \bigotimes_{e \in S(v)}  \rho_{e}^* \circ \phi_\BB (g_v) \right).
$$
\end{prop}
Finally, this leads us to consider the following orthonormal basis decomposition of the Hilbert space $L^2(\X/\G) \equiv L^2(\X)^\G$:
$$
L^2(\X/\G) \simeq \bigoplus_{\begin{smallmatrix}\{A_v,H_v\} \\ \{\rho_{e}, \BB_e\}  \end{smallmatrix}} \bigotimes_v \Inv(v,\rho),
$$
where $\Inv(v,\rho)$ are intertwining operators $\iota_v$ on each vertex $v$, {\it i.e.}
$$
\iota_v : \bigotimes_{e \in T(v)} \rho_e \to \bigotimes_{e \in S(v)} (\rho_{e})^{K_{\BB_e}} \circ \phi_\BB
$$
as representations of the group $U(A_v)$ (recall that $\rho_e$ is a representation of $U(A_{t(e)})$).

\begin{defn}
A {\rm gauge network} is the data $\{ \Gamma, (A_v,\lambda_v, H_v;\iota_v)_v, (\rho_{e},\BB_e)_e \}$ where
\begin{enumerate}
\item $\Gamma$ is a directed graph.
\item $(A_v,\lambda_v, H_v)$ is an object in the category $\cC_0^s$ for each vertex $v \in \Gamma^{(0)}$.
\item For each edge $e \in \Gamma^{(1)}$, $\rho_{e}$ is a representation of the group $G_e \equiv  \Aut_{\tilde A_{t(e)}}(H_{t(e)}) \times \U( \ker \lambda_{t(e)} )$.
\item For each edge $e \in \Gamma^{(1)}$, $\BB_e$ is a Bratteli diagram for $*$-algebra maps $A_{s(e)} \to A_{t(e)}$ with subdiagrams $\tilde \BB$ for $\tilde A_{s(e)} \to \tilde A_{t(e)}$, and $\BB_0$ for $A_{s(e)} \to \ker \lambda_{t(e)}$. 
\item For each vertex $v$, the $\iota_v$ are intertwiners for the group $\G_v \simeq U(A_v) \rtimes S(A_v)$:
$$
\iota_v : \rho_{e'_1} \otimes \cdots \otimes \rho_{e'_k} \to \rho_{e_1}^{K_{\BB_{e_1}}} \circ \phi_\BB \otimes \cdots \otimes \rho_{e_l}^{K_{\BB_{e_l}}} \circ \phi_\BB
$$
where $e_1' ,\ldots, e_k'$ are the incoming edges to $v$, $e_1, \ldots ,e_l$ are the outgoing edges from $v$ and the isotropy group $K_{\BB_e} = \U(\ker \lambda_{t(e)})_{\BB_{e0}}$.
\end{enumerate}
\end{defn}

\subsection{Examples of gauge networks}

We illustrate the above definition of gauge network by working out several examples. 

\subsubsection{Abelian spin networks}
\label{ex:network11}
\begin{figure}
\subfigure[$\Gamma$]{\includegraphics{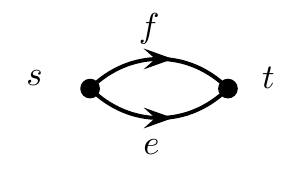}}
\subfigure[Bratteli diagram $\BB$]{\hspace{2cm}\includegraphics{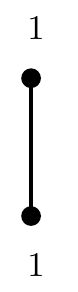}\hspace{2cm}}
\subfigure[A gauge network]{\includegraphics{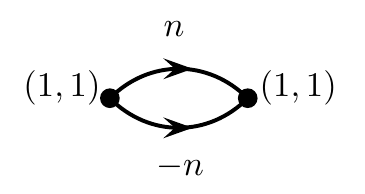}}
\caption{The graphs, Bratteli diagram and gauge network for the abelian case}
\label{fig:network11}
\end{figure}

Consider the directed graph $\Gamma$ of Figure \ref{fig:network11}, were we assign to both the source and target vertices $s$ and $t$ the pair $(\C,\C)$ of $*$-algebra and inner product space. A unital $*$-algebra map $\phi : \C \to  \C$ is necessarily given by $\phi( z ) = \tilde \phi(z)= z$, while the isometry $L: \C \to \C$ is given by $L(z) = u z$ for $u \in U(1)$. Consequently, $\X_e \simeq U(1)$ and $L^2(\X_e)$ decomposes into $U(1)$-representations:
$$
L^2(\X_e) \simeq \bigoplus_{n \in \Z} \C_{(n)} \otimes \C_{(-n)} 
$$
where $u \in U(1)$ acts on $\C_{(n)}$ via multiplication with the scalar $u^{n} $. A similar statement holds for $\X_f$. 

The corresponding gauge networks are equivalent to $U(1)$ spin networks on the graph $\Gamma$; they are given by assigning integers $n$ and $m$ to the edges $e$ and $f$, respectively, and intertwiners to the vertices
\begin{align*}
&\iota_{s} : \C \to  \C_{(n)} \otimes \C_{(m)} \qquad (U(1)\text{-representations})\\
&\iota_{t} : \C_{(n)} \otimes \C_{(m)} \to \C \qquad (U(1)\text{-representations})
\end{align*}
forcing $n=-m$. An example of a gauge network is thus given in Figure \ref{fig:network11}(c), where $(1,1)$ is shorthand for $(\C,\C)$.

\subsubsection{$U(N)$ spin networks}
\label{sect:spin-networks}
\begin{figure}
\includegraphics{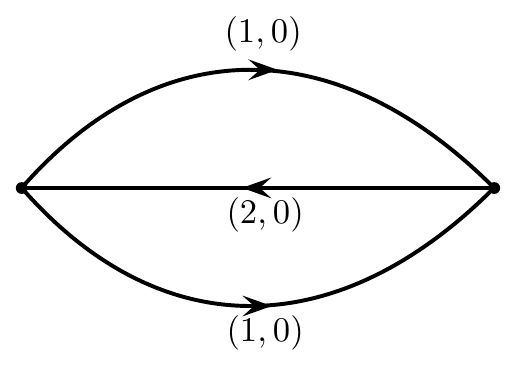}
\caption{A $U(2)$ spin network as a special case of a gauge network for the pair $(M_2(\C), \C^2)$ at both vertices. Here $(n, m)$ labels the highest weight representations of $U(2)$.}
\label{fig:network22}
\end{figure}

More generally, consider an arbitrary directed graph $\Gamma$ for which all algebras $A_v \simeq M_N(\C)$ and Hilbert spaces $H_v \simeq \C^N$. Our notion of gauge network for these choices of objects in $\cC_0$ reduces to the basis vectors called spin networks \cite{Bae94} for the Lie group $U(N)$. Indeed, there is only one Bratteli diagram $\BB = \tilde \BB$ ---depicted in Figure \ref{fig:Bratteli-ex0}--- and $G_e = \U(M_N(\C)) \equiv U(N)$ for all edges $e$. This leads to the data $(\Gamma, (\iota_v)_v , (\rho_e)_e)$ where the $\rho_e$ are representations of $U(N)$, while $\iota_v$ are intertwiners between the representations of $\G_v \simeq U(N)$ on the incoming edges and the outgoing edges to $v$: this is known in the literature as a {\it spin network}. An example is given in Figure \ref{fig:network22}.

\subsubsection{A gauge network associated to algebra maps $M_2(\C) \to M_4(\C)$}
\begin{figure}
\subfigure[$\Gamma$]{\includegraphics{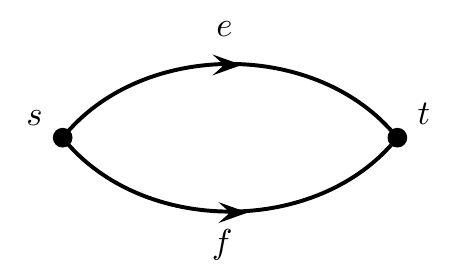}}
\subfigure[Bratteli diagram $\BB$]{\hspace{2cm}\includegraphics{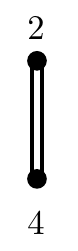}\hspace{2cm}}
\subfigure[A gauge network]{\includegraphics{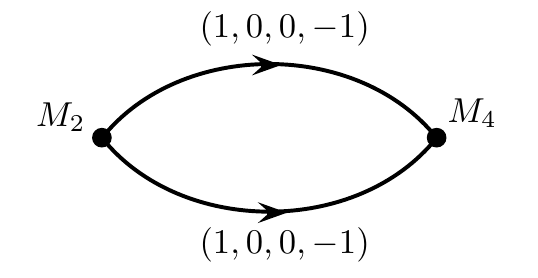}}
\caption{The graph $\Gamma$, Bratteli diagram and gauge network for algebra maps $M_2(\C) \to M_4(\C)$.}
\label{fig:network24}
\end{figure}

Now, we treat in some more detail a gauge network that is not a spin network. Consider the graph $\Gamma$ depicted in Figure \ref{fig:network24} and associate the following algebras
$$
A_{s} = M_2(\C); \qquad A_t = M_4(\C),
$$
and trivial Hilbert spaces $H_s = H_t = 0$ to the vertices. 

The Bratteli diagram in \ref{fig:network24}(b) indicates the algebra map 
$$
\phi_\BB : a \in M_2(\C) \mapsto  \begin{pmatrix} a& 0 \\ 0 & a \end{pmatrix} \in M_4(\C)  ,
$$
We also write $\phi_\BB(a) = 1_2 \otimes a  \in M_2(\C) \otimes M_2(\C) \simeq M_4(\C)$. Any other unital $*$-algebra map $\phi$ is related to $\phi_\BB$ by a change of basis. In other words, $\phi (\cdot) = u \phi_\BB(\cdot) u^*$ for some unitary $u \in U(4)$.

Consequently, the space $\X_e$ of unital $*$-algebra map $M_2 \to M_4$ is the orbit space of $\phi_\BB$ under the adjoint action of $u \in U(4)$. This means that $\X_e$ is the homogeneous space 
$$
\X_e \simeq U(4)/U(2)
$$
where $U(2)$ is the isotropy subgroup of $\phi_\BB$ ({\it i.e.} elements $u \in U(4)$ such that $u \phi_\BB(\cdot) u^* = \phi_\BB(\cdot)$, necessarily of the form $v \otimes 1_2$ for $v \in U(2)$). A similar statement hold for $\X_f$.

Next, there is an action of the gauge group $\G$ on $\X$, given in this case by a pair of unitaries in the source and target algebra: $u_1\in U(2)$ and $u_2 \in U(4)$.\footnote{Actually, we should have taken $u_1 \in PU(2)$ and $u_2 \in PU(4)$ in the projective unitary groups, as the centers of $U(2)$ and $U(4)$ act trivially on $A_s$ and $A_t$. However, since these centers also act trivially on $\X_e$ this is equivalent to taking $u_1 \in U(2)$ and $u_2 \in U(4)$.}
 They act on the map $\phi: M_2 \to M_4$ as
$$
\phi \mapsto u_1 \phi(u_2^* \cdot u_2) u_1^*.
$$
In terms of the unitary $u$ and $\phi_\BB$ we have
$$
u_1 \phi(u_2^* \cdot u_2) u_1^* = u_1 u \phi_\BB(u_2)^* \phi_\BB(\cdot) \phi_\BB(u_2) u  u_1^* 
$$
so that this action is implemented by the action of $(u_1,u_2) \in U(4) \times U(2)$ on $u \in U(4)$:
$$
u \mapsto u_1 u \phi_\BB(u_2^*).
$$

The crucial property is that the latter action induces an action on the quotient $\X$, since  $\phi_\BB(u_2^*)= 1_2 \otimes u_2^*$ commutes with the isotropy group, consisting of elements $v \otimes 1_2$ with $v \in U(2)$. We will indicate the two subgroups of $U(4)$ isomorphic to $U(2)$ suggestively as their matrix representation in $M_2 \otimes M_2$. In other words, we will write
\begin{align*}
1_2 \otimes U(2) &:= \left\{ \begin{pmatrix} u & 0 \\ 0 & u \end{pmatrix} : u \in U(2) \right\},\\
U(2) \otimes 1_2 &:= \left\{ \begin{pmatrix} u_{11} 1_2  & u_{12} 1_2  \\ u_{21} 1_2  & u_{22} 1_2 \end{pmatrix} : \begin{pmatrix} u_{11} & u_{12} \\ u_{21} & u_{22} \end{pmatrix} \in U(2) \right\}.
\end{align*}
Then, Proposition \ref{prop:gauge-network} becomes for this case:
\begin{prop}
With $\X= \X_e \times \X_f$, we have the corresponding Peter--Weyl decomposition:
$$
L^2(\X) \simeq \bigoplus_{\rho \in \widehat{U(4)}} \rho \otimes (\rho^*)^{U(2) \otimes 1}.
$$
With respect to this decomposition, the action of $\G = U(4) \times( 1 \otimes U(2))$ is given by 
$$
\rho(u_1) \otimes \rho^*(\phi_\BB(u_2)) \equiv\rho(u_1) \otimes \rho^*(1_2 \otimes u_2).
$$
\end{prop}

Recall that an irreducible finite-dimensional representation $\rho$ of $U(4)$ is characterized by a dominant weight $(m_1,m_2 , m_3, m_4)$, a quadruple of non-decreasing integers $m_1 \geq m_2 \geq m_3 \geq m_4$. More generally, a vector $v$ is said to have weight $(m_1,m_2 ,m_3, m_4)$ if it satisfies 
$$
\rho \begin{pmatrix} t_1 &0 & 0&0 \\ 0& t_2 & 0& 0 \\0 &0& t_3 &0 \\ 0&0&0& t_4 \end{pmatrix} v = t_1^{m_1}t_2^{m_2}t_3^{m_3}t_4^{m_4}v
$$
for the action of the diagonal elements in $U(4)$. We denote the $U(4)$-representation space for the dominant weight $(m_1,m_2,m_3,m_4)$ by $V_{(m_1,m_2,m_3,m_4)}$. It turns out that all weights in $V_{(m_1,m_2,m_3,m_4)}$ are the lattice points in the convex hull of all permutations of $(m_1,m_2,m_3,m_4)$. In what follows, we also need the multiplicities of these weight spaces; they can be obtained quite conveniently from so-called Gelfand-Tsetlin diagrams. We refer to the thesis \cite{Ras04} for an excellent review. 

\bigskip

Let us then consider the first non-trivial case of a gauge network for the above $\X_e\simeq \X_f \simeq  U(4)/U(2)$. It turns out to be given by the 15-dimensional $V_{(1,0,0,-1)}$, assigned to both edges $e$ and $f$. We first determine the invariant subspace $(V_{(1,0,0,-1)})^{U(2) \otimes 1}$, and then find two intertwiners 
\begin{equation}
\label{eq:intertwiners-U24}
\begin{aligned}
&\iota_s:  \C \to (V_{(1,0,0,-1)})^{U(2) \otimes 1} \otimes (V_{(1,0,0,-1)})^{U(2) \otimes 1}\qquad &(\text{as }1\otimes U(2)\text{-representations}),  \\
&\iota_t: V_{(1,0,0,-1)} \otimes V_{(1,0,0,-1)} \to \C \qquad & (\text{as }U(4)\text{-representations}). 
\end{aligned}
\end{equation}
The weight spaces in $V_{(1,0,0,-1)}$ are all one-dimensional (with weights given by the twelve permutations of $(1,0,0,-1)$, except that for weight $(0,0,0,0)$, which is 3-dimensional. Let us determine the reduced representation of $U(2) \otimes 1_2$ on $V_{(1,0,0,-1)}$. If the corresponding Lie algebra $\u(2) \otimes 1 \subset \u(4)$ is generated by $h_1, h_2, e$ and $f$, we have
\begin{gather*}
\rho|_{\u(2) \otimes 1} (h_1) = \rho(e_{11} + e_{22}) ;\qquad
\rho|_{\u(2) \otimes 1} (h_2) = \rho(e_{33} + e_{44}); \\
\rho|_{\u(2) \otimes 1} (e)= \rho(e_{13} + e_{24}); \qquad 
\rho|_{\u(2) \otimes 1} (f)= \rho(e_{31} + e_{42}).
\end{gather*}

Similarly, for the group $1_2 \otimes U(2)$ we have the corresponding Lie algebra $1 \otimes \u(2) \subset \u(4)$, say, generated by $h_1', h_2', e'$ and $f'$, we have
\begin{gather*}
\rho|_{1 \otimes \u(2)} (h_1') = \rho(e_{11} + e_{33}) ;\qquad
\rho|_{1 \otimes\u(2) } (h_2') = \rho(e_{22} + e_{44}); \\
\rho|_{1 \otimes\u(2) } (e')= \rho(e_{12} + e_{34}); \qquad 
\rho|_{1 \otimes\u(2) } (f')= \rho(e_{21} + e_{43}).
\end{gather*}

The decomposition of $V_{(1,0,0,-1)}$ into weight spaces and the relevant action of the two Lie algebras $\u(2) \otimes 1_2$ and $1_2 \otimes \u(2)$ can be summarized by Table \ref{table:U4-rep}. From this, we can easily read off the $U(2)\otimes 1_2$-invariant part, since it is given by the kernel of $\rho|_{\u(2) \otimes 1_2}(e)$ and $ \rho|_{\u(2) \otimes 1_2}(f)$. Now, note that on a $\u(2) \otimes 1_2$-weight zero vector $v$ we have 
$$
\langle \rho(e) v ,\rho(e) v \rangle =\langle \rho(f) v ,\rho(f) v \rangle
$$
so that on these weight zero spaces we have $\ker \rho(e) = \ker \rho(f)$. We conclude that the invariant subspace $(V_{(1,0,0,-1)})^{U(2)\otimes 1_2}$ is 3-dimensional. In fact, as a $1_2\otimes U(2)$-representation space:
\begin{equation}
\label{eq:U2-inv-U4}
(V_{(1,0,0,-1)})^{U(2) \otimes 1_2} \simeq V_{(1,-1)} ,
\end{equation}
where we have adopted the notation $V_{(n,m)}$ for $U(2)$-representation spaces with dominant weight $(n,m)$. 

\begin{table}
$$
\xymatrix{
(0,-1,1,0) &  {\begin{matrix} (1,-1,0,0) \\  \oplus \\(0,0,1,-1) \end{matrix}} \ar[l]_{\rho(f)} \ar[r]^{\rho(e)} \ar@/^/[d]^{\rho(f')} 
& (1,0,0,-1)
\\
{\begin{matrix} (-1,0,1,0)\\ \oplus \\(0,-1,0,1)\end{matrix}} &(0,0,0,0)  \ar@/^/[u]^{\rho(e')}\ar@/^/[d]^{\rho(f')} \ar[l]_{\rho(f)} \ar[r]^{\rho(e)} & {\begin{matrix} (1,0,-1,0) \\\oplus\\ (0,1,0,-1)\end{matrix}}
\\
(0,-1,1,0) & {\begin{matrix} (1,-1,0,0)\\ \oplus\\ (0,0,1,-1) \end{matrix}}\ar[l]_{\rho(f)} \ar[r]^{\rho(e)}
\ar@/^/[u]^{\rho(e')} & (1,0,0,-1)
}
$$
\caption{The decomposition of $V_{(1,0,0,-1)}$ into weight spaces and the action of $\u(2) \otimes 1_2$ and $1_2 \otimes \u(2)$.}
\label{table:U4-rep}
\end{table}

Next, we determine two intertwiners $\iota_s$ and $\iota_t$ as in Eq. \eqref{eq:intertwiners-U24}. As usual, we have tensor product decompositions,
\begin{align*}
(V_{(1,0,0,-1)})^{U(2) \otimes 1_2} \otimes (V_{(1,0,0,-1)})^{U(2) \otimes 1_2}
& \simeq V_{(1,-1)} \otimes V_{(1,-1)} 
\simeq V_{(2,-2)} \oplus V_{(1,-1)} \oplus V_{(0,0)}
\end{align*}
as $1_2 \otimes U(2)$-representations, and similarly,
 \begin{align*}
V_{(1,0,0,-1)} \otimes V_{(1,0,0,-1)}
& \simeq V_{(2,0,0,-2)} \oplus \cdots \oplus V_{(0,0,0,0)}
\end{align*}
as $U(4)$-representations. 

In particular, both tensor products contain the trivial representation, $V_{(0,0)}$ for $1_2 \otimes U(2)$ and $V_{(0,0,0,0)}$ for $U(4)$, respectively. This surely allows for intertwiners $\iota_s$ and $\iota_t$ and to the gauge network as depicted in Figure \ref{fig:network24}(c).

\subsubsection{A gauge network associated to algebra maps $M_2(\C) \to M_2(\C) \to M_4(\C)$}

\begin{figure}
\includegraphics{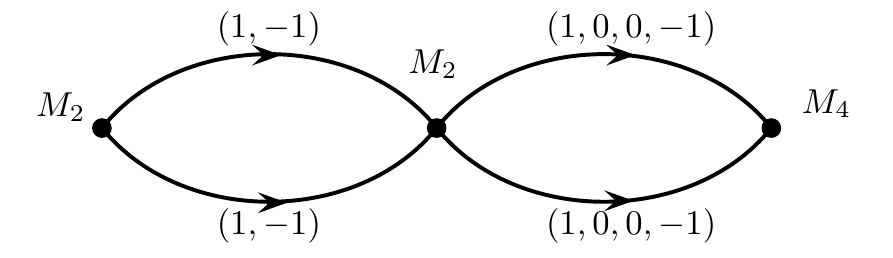}
\caption{A gauge network for the algebra maps $M_2(\C) \to M_2(\C) \to M_4(\C)$.}
\label{fig:network224}
\end{figure}

We consider the graph $\Gamma$ depicted in Figure \ref{fig:network224} with the following algebras associated to the vertices $s, v$ and $t$:
$$
A_{s} = M_2(\C); \qquad A_v = M_2(\C) ; \qquad A_{t} = M_4(\C),
$$
and, again trivial Hilbert spaces $H_{s} = H_v = H_{t} = 0$ to the vertices. 

Arguing as in the previous example ---noting in addition that the space of $*$-algebra maps $M_2(\C) \to M_2(\C)$ is given by $U(2)/U(1)$--- the relevant configuration space is
$$
\X \simeq \left(\frac{ U(2)  \times U(2)}{U(1) \times U(1) } \right) \times \left( \frac{ U(4)  \times U(4)}{U(2) \times U(2) }\right)
$$
with gauge group 
$$
\G \simeq U(2) \times U(2) \times U(4).
$$
We have ordered the product of homogeneous spaces in $\X$ suggestively, so as to have
$$
\X/\G \simeq  U(2) \backslash \left(\frac{ U(2)  \times U(2)}{U(1) \times U(1) } \right) \times_{U(2)} \left( \frac{ U(4)  \times U(4)}{U(2) \times U(2) }\right)/U(2)
$$
Again, actually the gauge group is the projective group $PU(2) \times PU(2) \times PU(4)$ but since the center $U(1) \times U(1) \times U(1)$ acts trivially on $\X$ we can just as well consider the above action of $\G$ on $\X$. 

An example of a gauge network is given as in Figure \ref{fig:network224}, labelling the edges in $\Gamma$ by representation spaces $V_{(1,-1)}$ of $U(2)$ and $V_{(1,0,0,-1)}$ of $U(4)$. For the invariant subspaces, we have
\begin{align*}
&(V_{(1,-1)})^{U(1)} \simeq  V_{(1,-1)} \qquad &(\text{as } U(2)\text{-representations});\\
&(V_{(1,0,0,-1)})^{U(2) \otimes 1_2} \simeq V_{(1,-1)} \qquad &(\text{as } 1_2 \otimes U(2)\text{-representations}),
\end{align*}
as we already established in Equation \eqref{eq:U2-inv-U4}. 

Next, we need to find intertwiners at the three vertices of the following form:
\begin{equation*}
\label{eq:intertwiners-U224}
\begin{aligned}
&\iota_{s}:  \C \to (V_{(1,-1)})^{U(1)} \otimes (V_{(1,-1)})^{U(1)} \quad &(\text{for }U(2)),  \\
&\iota_{v}: V_{(1,-1)} \otimes V_{(1,-1)} \to 
(V_{(1,0,0,-1)})^{U(2) \otimes 1_2} \otimes (V_{(1,0,0,-1)})^{U(2) \otimes 1_2}
 \quad & (\text{for }1_2 \otimes U(2)), \\
&\iota_{t}: V_{(1,0,0,-1)} \otimes V_{(1,0,0,-1)} \to \C \quad & (\text{for }U(4)). 
\end{aligned}
\end{equation*}
In other words, we need to find intertwiners
\begin{equation*}
\begin{aligned}
&\iota_{s}:  \C \to V_{(2,-2)} \oplus V_{(1,-1)} \oplus V_{(0,0)} \quad &(\text{for }U(2)),  \\
&\iota_{v}: V_{(2,-2)} \oplus V_{(1,-1)} \oplus V_{(0,0)} \to
V_{(2,-2)} \oplus V_{(1,-1)} \oplus V_{(0,0)}
 \quad & (\text{for }U(2)), \\
&\iota_{t}: V_{(1,0,0,-1)} \otimes V_{(1,0,0,-1)} \to \C \quad & (\text{for }U(4)). 
\end{aligned}
\end{equation*}
which, as one can readily check, is indeed possible.

\section{Correspondences between gauge networks}\label{corrSec}
We introduce the notion of correspondence between gauge networks and consider them as morphisms in a category of gauge networks. We motivate our construction by starting to consider morphisms between elements in the space $\X$. 

Given two elements $\pi$ and $\pi'$ in $\X$, on each vertex $v \in \Gamma^{(0)}$ we have, say, algebras $A_v$ and $A_v'$, respectively. A morphism between $\pi$ and $\pi'$ should at least be a morphism between these algebras and the natural candidate to consider is an $A_v-A_v'$-bimodule $E_v$, which we denote by the diagram
$$
\xymatrix{&E_v \ar[dr] \ar[dl] \\ A_v & &A_v'}
$$ 
Note that the arrows do not represent maps, merely the interpretation of $E_v$ as a correspondence between $A_v$ and $A_v'$.
We also require the vector spaces $H_v$ and $H_v'$ to be related via
$$
H_v \simeq E \otimes_{A_v'} H_v',
$$
compatibly with the action of $A_v$. Finally, along the edges of $\Gamma$ one should have a map $T_e: E_{s(e)} \to E_{t(e)}$, compatibly with the algebra maps $\phi_e$ and $\phi_e'$:
$$
T_e(a\eta b) = \phi_e(a) T_e(\eta) \phi'_e(b); \qquad (a \in A_{s(e)}, \eta \in E_{s(e)}, b \in A'_{s(e)}).
$$
This compatibility is conveniently denoted by a diagram:
$$
\xymatrix{&E_{s(e)} \ar[dr] \ar[dl] \ar[dd]_{T_e} \\ A_v\ar[dd]_{\phi_e} & &A'_v \ar[dd]_{\phi_e'} \\ & E_{t(e)} \ar[dr] \ar[dl] \\ A_{t(e)} &&A'_{t(e)}}
$$ 
If we dualize this construction by going to $L^2$-spaces as above, and take the $\G$-invariance into account, we arrive at the following notion of correspondence between gauge networks:
\begin{defn}
Let $\psi$ and $\psi'$ be two gauge networks on the same graph $\Gamma \subset M$:
\begin{gather*}
\psi = (\Gamma, (A_v,H_v,\iota_v)_v, (\rho_{e},\BB_e)_e), \qquad
\psi' = (\Gamma, (A_v',H_v',\iota_v)_v, (\rho'_{e},\BB'_e)_e).
\end{gather*}
A {\rm correspondence} $\Psi$ between $\psi$ and $\psi'$ is the data $\{ \Gamma, ({}_{A_v} E_{A_v'},\iota_v \otimes \iota_v')_v, (\rho_{e} \otimes \rho'_{e},\BB_e \times \BB_e')_e \}$ where in addition to the above, ${}_{A_v} E_{A_v'}$ is an $A_v-A_v'$-bimodule. 
\end{defn}
If needed, we will use the notation ${}_\psi \Psi_{\psi'}$ to indicate the source and target of the morphism. We can compose two correspondences $\Psi_1$ and $\Psi_2$ when the target of $\Psi_1$ coincides with the source of $\Psi_2$. That is, if 
\begin{align*}
\Psi_1 &= \{ \Gamma, ({}_{A_v} E_{A_v'},\iota_v \otimes \iota_v')_v, (\rho_{e} \otimes \rho'_{e}, \BB_e \times \BB_e')_e \}\\
\Psi_2 &= \{ \Gamma, ({}_{A_v'} F_{A_v''},\iota_v' \otimes \iota_v'')_v, (\rho'_{e} \otimes \rho''_{e}, \BB_e' \times \BB_e'')_e \}
\end{align*}
we define
$$
\Psi_1 \circ \Psi_2 = \{ \Gamma, ({}_{A_v} E \otimes_{A_v'} F_{A_v''} ,\iota_v \otimes \iota_v'')_v, (\rho_{e} \otimes \rho''_{e}, \BB_e \times \BB_e'')_e \}
$$ 

We denote by $\S$ the category of gauge networks with as morphisms the above correspondences. Associated to it, we can form the algebra generated by the morphisms in $\S$, denoted by $\C[\S]$. It consists of elements of the form
$$
a = \sum_{\Psi} a_\Psi \Psi
$$
where only finitely many $a_\Psi \neq 0$ and the sum ranges over all gauge network correspondences. The composition of morphisms translates into a convolution product in $\C[\S]$:
$$
(a \ast b)_\Psi = \sum_{\Psi= \Psi_1 \circ \Psi_2} a_{\Psi_1} b_{\Psi_2}.
$$

\subsection{$C^*$-algebra on gauge network correspondences}
The above algebraic construction can be extended to a $C^*$-algebraic setting by letting $\C[\S]$ act on $L^2(\X/\G)$. We introduce a representation $\pi_\S$ of $\C[\S]$ by setting
$$
\pi_\S (a) \psi = \sum_{{}_{\psi'} \Psi_{\psi}} a_\Psi \psi'
$$
on the gauge networks $\psi$ (which are basis vectors of $L^2(\X/\G)$. One readily checks that $\pi_\S(a \ast b) = \pi_\S(a) \pi_\S(b)$, as required. The $C^*$-algebraic completion of $\C[\S]$ in this representation will be denoted by $C^*(\S)$.

\begin{rem}
There is also another Hilbert space on which $\C[\S]$ can act, naturally associated to the category $\S$ of gauge network correspondences.
If we fix a gauge network $\psi_0$ we can restrict $\S$ to morphisms to $\psi_0$ by setting
$$
\S_{\psi_0} := \{ \Psi \in \Hom_\S(\psi_0, \psi) \text{ for some } \psi \}.
$$
The Hilbert space $l^2(\S_{\psi_0})$ carries a representation $\pi_{\psi_0}$ of $\C[\S]$:
$$
\pi_{\psi_0} (a) \xi(\Psi) = \sum_{\Psi=\Psi_1 \circ \Psi_2} a_{\Psi_1} \xi(\Psi_2); \qquad ( \xi \in l^2(\S_{\psi_0}))
$$
\end{rem}

\subsection{Time evolution}

Recall the form \eqref{eq:L2-A} of $L^2(\X)$. We introduce a Hamiltonian operator $\bH$ on $L^2(\X)$ as the sum of the invariant Laplacians on the homogeneous spaces. In fact, the quadratic Casimir operators of the Lie groups $\U(A_{t(e)})$ are $\U(A_{t(e)})$-bi-invariant, so that on each $L^2(\U(A_{t(e)})/\U(\lambda_{t(e)})_{\BB_{e0}})$ we have an induced Laplacian operator $C^2_{\U(A_{t(e)})} $. 
\begin{prop}
\label{prop:hamiltonian}
\begin{enumerate}
\item 
After choosing pairs $\{A_v,H_v\}$ at each vertex $v \in \Gamma^{(0)}$ and Bratteli diagrams $\BB_e$ at each edge $e \in \Gamma^{(1)}$, the tensor product
$$
\sum_e \left(1 \otimes \cdots \otimes C^2_{\U(A_{t(e)})} \otimes \cdots \otimes 1 \right) 
$$
of quadratic Casimirs is an essentially self-adjoint operator on the finite tensor product $\bigotimes_e L^2( \U(A_{t(e)}) / \U(\lambda_{t(e)})_{\BB_{e0}})$. 
\item The sum of the operators defined in (1) is an essentially self-adjoint operator $\bH$ on $L^2(\X)\simeq \bigoplus_{\{A_v,H_v\}} \bigoplus_{\{\BB_e\}} \bigotimes_e L^2( \U(A_{t(e)})/\U(\lambda_{t(e)})_{\BB_{e0}})$. 
\end{enumerate}
\end{prop}
\proof
The first claim follows from \cite[Theorem VIII.33]{RS72}. For the second, if $\{T_k\}_{k=1}^\infty$ is an infinite series of essentially self-adjoint operators on Hilbert spaces $\H_k$ (say, each with core $D_k \subset \H_k$) then $\sum_k T_k$ is essentially self-adjoint on $\bigoplus_k \H_k$ with core $D= \bigoplus_{k=1}^\infty D_k$ (finitely many combinations). 
\endproof

Moreover, by the same $\U(A_{t(e)})$-invariance that was noted before, the operator $\bH$ commutes with the action of $\G$ on $L^2(\X)$. Hence, it makes sense to consider the induced operator on the $\G$-invariant subspace $L^2(\X/\G) \subset L^2(\X)$. We use the same notation for the induced essentially self-adjoint operator:
\begin{equation}
\label{eq:hamiltonian}
\bH: L^2(\X/\G) \to L^2(\X/\G).
\end{equation}

The interesting property of the operator $\bH$ is that it induces a time evolution on the $C^*$-algebra $C^*(\S)$ introduced above. 
\begin{prop}
There is a one-parameter group $(\sigma_t)_t$ of automorphisms of $C^*(\S)$ induced by $\bH$:
$$
\pi_\S (\sigma_t(a)) = e^{i t \bH} \pi_\S(a) e^{-i t \bH} ; \qquad (a \in C^*(\S)).
$$
\end{prop}
\proof
In terms of the finitely supported functions $a \in \C[\S]$ we have
$$
(\sigma_t(a))_\Psi = e^{i t (\psi ,\bH \psi) - it (\psi', \bH \psi')} a_\Psi
$$
if $\Psi = {}_\psi \Psi_{\psi'}$. Being a multiplication of $a_\Psi$ by a phase factor, $\sigma_t(a) \in \C[\S]$ and, moreover, it is continuous in $t$ with respect to the operator norm when acting on $L^2(\X/\G)$. Hence, $\sigma_t$ extends to an automorphism of the $C^*$-algebraic completion.
\endproof

\section{The spectral action and lattice field theory}\label{latticeSec}
The above discussion on spin networks involved an abstract graph; we will now connect to a background geometry by embedding $\Gamma$ in a smooth spin manifold $M$. 
The above quiver representations gives rise to a twisted Dirac operator on $\Gamma$. We will show that the corresponding spectral action reduces to the Wilson action of lattice gauge theory.

\subsection{The spin geometry of $\Gamma$}
Suppose that $\Gamma$ is embedded in a Riemannian spin manifold $M$. Then, we can pullback some of the spin geometry on $\Gamma$. Let $\S$ be the typical fiber of the spinor bundle on $M$. The space of $L^2$-spinors on $\Gamma$ will then be $\S^{\Gamma^{(0)}}$. A `Dirac operator' can be defined using the holonomy $\Hol(e, \nabla^S)$ of the spin connection along the edges $e$ of $\Gamma$:
$$
(D_{\Gamma}\psi)_v = \sum_{t(e)=v} \frac{1}{2 l_e} \gamma_e  \Hol(e, \nabla^S) \psi_{s(e)} + \sum_{s(e)=v} \frac{1}{2 l_{\bar e}} \gamma_{\bar e} \Hol(\bar e, \nabla^S) \psi_{t(e)}; \qquad (\psi \in \S^{\Gamma^{(0)}}),
$$
where $l_e$ is the geodesic length of the (embedded) edge $e$ in $M$ and $\bar e$ is the (embedded) edge $e$ with reverse orientation. The gamma matrices $\gamma_e$ are defined as follows. At a vertex $v$, consider the span $E$ of the vectors $\dot e_i$ in $T_v M$ defined by the outgoing edges $e_i$ at $v$. Let $\partial_\mu$ be an orthonormal basis of $E$, related to $\dot e_i$ via
$$
\dot e_i = X_{e_i}^\mu \partial_\mu.
$$
Then, we define covectors $\theta^{e^i}$ colinear with $\dot e_i$ such that 
$$
\sum_i \theta^{e^i} X_{e_i}^\mu = dx^\mu.
$$
We set $\gamma_e := i c(\theta^e)$, in terms of Clifford multiplication by the covector $\theta^e$. 
The crucial property is that for a one-form $\omega \in \Omega^1(M)$:
$$
\sum_{e \in S(v)} \gamma_e \omega_e = \sum c(\theta^e) X_e^\mu \omega_\mu = \gamma^\mu \omega_\mu
$$
where $\omega_e \equiv \langle \omega, \dot e \rangle$ and $\omega_\mu = \langle \omega,\partial_\mu \rangle$. 

One checks from $\gamma_{\bar e} = \gamma_e^*$ and $\Hol(e,\nabla^S)^* = \Hol(\bar e, \nabla^S)$ that $D_{\Gamma}$ is symmetric. In fact, the triple
$$
(\C^{\Gamma^{(0)}} , \S^{\Gamma^{(0)}} , D_\Gamma)
$$
is a finite spectral triple. 

\subsubsection{Continuous limit of the Dirac operator}
We now let the `lattice spacing' $l_e$ go to zero, further assuming that the above subspace $E$ spanned by the edges at each vertex $v$ actually spans $T_v M$. Moreover, we assume $l_e = l$ is the same for all edges, and suppose we are on a square lattice. The key property of the holonomy is that, at first order in $l$ we have
$$
\Hol(e,\nabla^S) = \cP e^{\int_e \omega\cdot dx} \sim 1+ l \omega_e(s(e)) + \cO(l^2)
$$
in terms of the spin connection one-form $\omega$. Here $\omega_e(v)$ means the value of the pairing between the one-form $\omega$ and the vector $\dot e$ at the vertex $v$. So, up to terms of order $l$, we have
$$
(D_{\Gamma} \psi)_v = \sum_{v_1,v_2} \frac{1}{2l} \gamma_e ( \psi_{v_1} - \psi_{v_2}) + \frac{1}{2} \gamma_e \omega_e(v) (\psi_{v_1} + \psi_{v_2}) + \cO(l).
$$
where the sum is over all colinear edges that connect at $v$, indicated by the connecting vertices $v_1$ and $v_2$, as in:
$$
\xymatrix{&v_1 \ar[r]_{e'} &v \ar[r]_{e} &v_2}
$$
Indeed, in this case, at $v$ we have $\dot e' = - \dot{\bar {e}}$, so that $\gamma_{\bar {e'}} = - \gamma_{e}$ as well as $\omega_{\bar{e'}} = - \omega_{e}$. We conclude that, at least formally, in the limit that $l \to 0$:
\begin{equation}
\label{eq:dirac-limit}
(D_{\Gamma} \psi)_v \to \gamma^\mu( \partial_\mu  + \omega_\mu )\psi(v)
\end{equation}
which we recognize as the Dirac operator on $M$, evaluated on a spinor $\psi$ at $v \in M$.

Since we only consider this as a motivation for our construction, we will not dwell further on the technical details of this derivation.

\subsection{Twisted Dirac operator and lattice gauge fields}
Suppose that in addition to the spin geometry on $M$ we are given representations of the quiver $\Gamma$ in the category $\cC$. Thus, along with the spin connection on each edge, we have a linear map $L_e: H_{s(e)} \to H_{t(e)}$. Moreover, on each vertex $v$ we have a finite Dirac operator $D_v$. Hence, introducing a Hilbert space  $\S \otimes (\bigoplus_v  H_v)$, we define a `twisted Dirac operator' by
$$
(D_{\Gamma,L}\psi)_v = \sum_{t(e)=v} \frac{1}{2 l_e} \gamma_e  \left(\Hol(e, \nabla^S) \otimes L_e \right) \psi_{s(e)} + \sum_{s(e)=v} \frac{1}{2 l_{\bar e}} \gamma_{\bar e} \left( \Hol(\bar e, \nabla^S) \otimes L_{\bar e} \right) \psi_{t(e)}
+ \gamma D_v \psi_v,
$$
where $L_{\bar e}$ is defined as the hermitian conjugate of the isometry $L_{e}$. Also, $\gamma$ denotes the grading on the spinor bundle on $M$, assuming $M$ is even dimensional. 

\begin{rem}
In \cite{Kra96} a diagrammatic classification of finite spectral triples was given, with vertices corresponding to representations of the finite-dimensional $*$-algebra, and edges corresponding to a non-zero finite Dirac operator between them. This is very similar to the above definition of a Dirac operator on the quiver $\Gamma$, allowing for a speculative but intriguing picture in which one cannot distinguish the (discretized) spin geometry of $M$ from the finite noncommutative geometry $(A_v,H_v,D_v)$ at each vertex.
\end{rem}

\begin{prop}
\label{prop:dirac-gauge}
The gauge group $\G = \coprod_{\{A_v,H_v\}} \prod_v \G_v$ acts on $\S \otimes (\bigoplus_v  H_v)$ by unitary operators $U(g)$. Explicitly, with $g_v \in \U(A_v)$:
$$
(U(g)\psi)_v = \lambda_v( g_v) \psi_v
$$
Moreover, for vanishing $D_v$, the twisted Dirac operator $D_{\Gamma,L}$ satisfies
$$
D_{\Gamma, g(L)} = U(g) \circ D_{\Gamma,L} \circ U(g)^*
$$
where $(g(L))_e = \lambda_{t(e)}( g_{t(e)}) L_e \lambda_{s(e)}(g_{s(e)})^*$ is the action of Proposition \ref{prop:gauge-group}. If $D_v \neq 0$ then we also have $D_v \mapsto \lambda_v(g_v) D_v \lambda_v(g_v)^*$.
\end{prop}

A link to classical geometry of and gauge theory on $M$ can be established as follows. If, in addition to the assumptions made in the previous subsection, we suppose that the pairs $(A_v,H_v)= (M_N(\C), \C^N)$, for all vertices $v$, then a non-zero morphism $(\phi,L)$ is a unitary map in $U(N)$. If we think of it as the holonomy of some gauge connection one-form $A_\mu$ we can derive, up to first order in $l$, that the above twisted Dirac operator on $\Gamma$ reduces to the Dirac operator on $M$, twisted by the gauge field $A_\mu$.

More interesting is to consider the twisted Dirac operator on $\Gamma$ in its own right. Since our construction is finite-dimensional, it is obvious that the triple
$$
\left( \bigoplus A_v, \S \otimes (\bigoplus_v  H_v), D_{\Gamma,L}\right)
$$
is a {\it spectral triple}. Any such triple gives rise to a unitary gauge group consisting of the unitaries $\U(\bigoplus A_v)$ \cite{CM07}. Thus, in the case of faithful $*$-algebra representations $\lambda_v$ on $H_v$, this group coincides with the gauge group $\G$. 

A natural gauge invariant functional associated to a spectral triple is the {\it spectral action} \cite{CC96} on it, which in our case is
\begin{equation}
\label{eq:sa}
S[\{L_e\}, \{D_v\}]= \tr f( D_{\Gamma,L})
\end{equation}
for some function $f$ on the real line. The {\it fermionic action} is defined by
$$
S_F[\{ \psi_v\},\{L_e\}, \{D_v\}] = \left \langle \psi, D_{\Gamma,L} \psi \right \rangle
$$

\begin{prop}
\label{prop:sa-gauge}
Both the spectral action and the fermionic action are invariant under the action of the gauge group: $L_e \mapsto g_{t(e)} L_e g_{s(e)}^*$ and $D_v \mapsto \lambda_v (g_{v}) D_v \lambda_v(g_{v})^*$ with $g_v \in \G_v$.
\end{prop}
\proof
This follows directly from Proposition \ref{prop:dirac-gauge}.
\endproof

In other words, the spectral and fermionic action define functions on the configuration space $\X/\G$. Moreover, they act on $L^2(\X/\G)$ by bounded operators. As such, the spectral action can be added as an interaction term to the Hamiltonian operator $\bH$ defined in Equation \eqref{eq:hamiltonian}. This allows to put the spectral action in the same position as the Wilson action in lattice gauge theory, a fact that we will now further work out.

\subsubsection{Lattice gauge fields}
We first assume that $\Gamma$ is a four-dimensional square lattice, so $M= \R^4$.
We will show that if we take $f(x)= x^4$ for the spectral action, then $S[L]$ gives rise to the Wilson action in lattice gauge theory. Moreover, the variables $D_v$ at the vertices can be interpreted as scalar fields and the spectral action reproduces the Higgs-field lattice system \cite{LRV81,DJK84} for a Higgs field in the adjoint representation. Finally, the action $S_F$ gives rise to the usual action for fermions on a lattice, coupled to gauge and Higgs field. 

It is convenient to take as in \cite{CC96} a cutoff $ \Lambda \propto l^{-1}$ and compute
$$
S_\Lambda [\{L_e\}, \{D_v\}]:= \tr f( D_{\Gamma,L}/ \Lambda ) \equiv l^4 \tr (D_{\Gamma,L})^4.
$$
We obtain the following result.

\begin{thm}\label{thm:sa}
The spectral action \eqref{eq:sa} is given by
\begin{equation}\label{SLambda}
\begin{array}{rl}
S_\Lambda[\{L_e\},\{D_v\}] &= - \frac1 {4} \sum_{\partial p = e_4 \cdots e_1} 
 \left( \tr \left( L_{\bar e_4} L_{\bar e_3} L_{e_2} L_{e_1} \right)
+ \tr \left( L_{\bar e_1} L_{\bar e_2} L_{e_3} L_{e_4} \right)\right) 
+ \text{const.} \\[3mm]
& \quad+ \sum_v  l^4 \tr D_v^4 + 4 l^2 \sum_e \left( \tr D_{s(e)}^2 + \tr D_{t(e)}^2 - \tr L_e^* D_{t(e)} L_e D_v \right)
\end{array}
\end{equation}

\end{thm}

\proof
Let us first consider the case that $D_v = 0$. Since the lattice is a square lattice $\Z^4 \subset M=\R^4$, the above form for $f$ selects precisely the trace over maps $L_e$ along edges that form a cycle of length 4 in $\Gamma$, thereby including edges $\bar e$ with reverse orientation (corresponding to the map $L_{\bar e}$ in the definition of $D_{\Gamma,L}$. Such cycles are of either one of the following forms: 
\begin{enumerate}
\item $c = \bar e_4 \bar e_3 e_2 e_1$ or $\bar e_1 \bar e_2 e_3 e_4$ as in Figure \ref{fig:plaquette}; here the vertex $v$ can also appear at the other three corners of the plaquette.
\label{item:cycle1}
\begin{figure}
\includegraphics{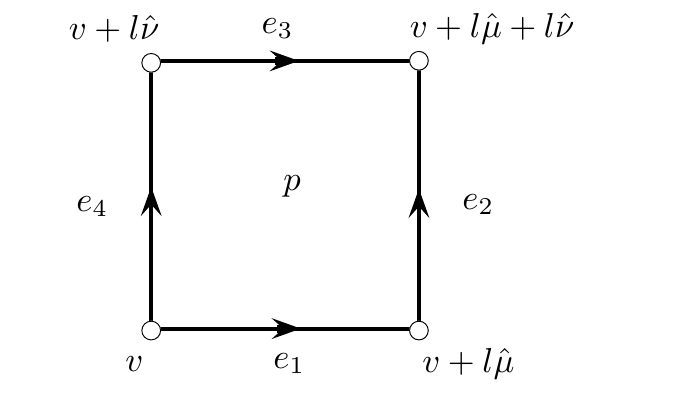}
\caption{A plaquette in the lattice; the vectors $\hat \mu$ and $\hat \nu$ correspond to the edges along the plaquette $p$.}
\label{fig:plaquette}
\end{figure}

\item  $c = \bar e_1 \bar e_2 e_2 e_1$ as in Figure \ref{fig:backforth}; the vertex $v$ can also be the middle vertex, in which case $e_1 \bar e_1 \bar e_2 e_2$ and $\bar e_2 e_2 e_1 \bar e_1 $ are possible cycles based at $v$.
\label{item:cycle2}
\begin{figure}
\includegraphics{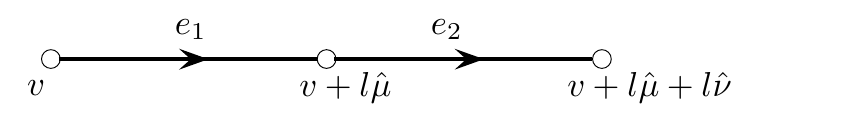}
\caption{Two edges in the lattice; the vectors $\hat \mu$ and $\hat \nu$ correspond to the edges $e_1$ and $e_2$.}
\label{fig:backforth}
\end{figure}

\item the trivial cycle $c = \bar e e \bar e e $ with $s(e) = v$ (cf. Figure \ref{fig:higgs-edge}); the vertex $v$ can also be the right vertex (i.e. $v=r(s)$, in which case $e \bar e e \bar e$ is a cycle based at $v$.
\label{item:cycle3}
\begin{figure}
\includegraphics{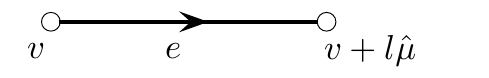}
\caption{A single edge $e$ in the lattice along the direction $\hat \mu$.}
\label{fig:higgs-edge}
\end{figure}
\end{enumerate}

If we also take into account that in the flat case the holonomy of the spin connection is trivial, and $\gamma_{e_1} = - \gamma_{\bar e_3} = \gamma_\mu$, $\gamma_{e_2} = - \gamma_{\bar e_4} = \gamma_\nu$, the trace in $S_\Lambda$ becomes 
\begin{align}\label{SLambda2}
S_\Lambda[\{L_e\}] &= 4 l^4 \sum_{\partial p = \bar e_4 \bar e_3 e_2 e_1} \frac{1}{(2l)^4} \tr (\gamma_\nu \gamma_\mu)^2
\left(\tr \left( L_{\bar e_4} L_{\bar e_3} L_{e_2} L_{e_1} \right)
+ \tr \left( L_{\bar e_1} L_{\bar e_2} L_{e_3} L_{e_4} \right)\right) + \text{const.}
\nn \\
&= 
- \frac14 \sum_{\partial p =\bar e_4  \bar e_3 e_2 e_1} \left(\tr \left( L_{\bar e_4} L_{\bar e_3} L_{e_2} L_{e_1} \right)
+ \tr \left( L_{\bar e_1} L_{\bar e_2} L_{e_3} L_{e_4} \right)\right) + \text{const.}
\end{align}
The factor 4 comes from the 4 possible choices for the vertex $v$ at the corners of a plaquette. The constant term comes from the contributions of cycles of type \eqref{item:cycle2} and \eqref{item:cycle3}, typically of the form $\tr L_{\bar e_1} L_{\bar e_2} L_{e_2} L_{e_1} = \tr (1)$. In what follows, we will ignore this constant term. 

In the general case, there are additional contributions from the action of $D_v$. Namely, the trace in $S_\Lambda$ now also involves a sum over cycles of length 2, given for each edge $e$ by combinations of $L_{\bar e}, L_e$ and $D_{s(e)}$ or $D_{t(e)}$. More precisely, for an edge $e$ the additional contributions to $S_\Lambda$ are
\begin{align*}
 l^4 \frac{1}{(2l)^2}\tr (\gamma_{\bar e} \gamma \gamma_e \gamma) \tr (L_e^* D_{t(e)} L_e D_{s(e)})
& = - l^2 \tr (L_e^* D_{t(e)} L_e D_{s(e)}),\\
 l^4 \frac{1}{(2l)^2}\tr (\gamma \gamma_{\bar e} \gamma \gamma_e ) \tr (L_e^* D_{t(e)} L_e D_{s(e)})& = - l^2 \tr ( D_{s(e)} L_e^* D_{t(e)} L_e),\\
\end{align*}
and also 
\begin{align*}
l^4 \frac{1}{(2l)^2} \tr (\gamma_{\bar e}  \gamma_e  \gamma^2) \tr (L_e^* L_e  D_{s(e)}^2) &= l^2  \tr ( D_{s(e)}^2),
\\
 l^4 \frac{1}{(2l)^2}\tr (\gamma^2 \gamma_{\bar e}  \gamma_e  ) \tr (D_{s(e)} L_e^* L_e D_{s(e)}  ) &=l^4   \tr ( D_{s(e)}^2),\\
 l^4 \frac{1}{(2l)^2}\tr (\gamma\gamma_{\bar e}  \gamma_e \gamma  ) \tr (D_{s(e)}^2 L_e^* L_e  ) &=l^4   \tr ( D_{s(e)}^2),\\
l^4 \frac{1}{(2l)^2} \tr (\gamma_{\bar e} \gamma^2 \gamma_e) \tr (L_e^* D_{t(e)}^2 L_e ) &= l^4  \tr ( D_{t(e)}^2)
\end{align*}
Taking into account all such possible cycles and also the single contribution from applying $D_v^4$ at each vertex, we arrive at Eq. \eqref{SLambda}.
\endproof

The expression \eqref{SLambda2} is very similar to the Wilson action (cf. \cite{Cre83, Mak02}). Below, we will show that if $(A_v,H_v) = (M_N(\C), \C^N)$ for all vertices $v$, it indeed induces the $U(N)$ Yang--Mills action when taking the 
continuum limit $l \to 0$.

In the last term in \eqref{SLambda}, one recognizes the gauge Higgs-field action on a lattice \cite{LRV81,DJK84}. In fact, the above action gives rise to the action for the Yang--Mills--Higgs system when taking the continuum limit $l \to 0$, as we will now show explicitly.

\subsubsection{Continuum limit of lattice gauge theory}

We recall that in the continuum limit $l \to 0$ the Wilson action reduces to the Yang--Mills action. This follows upon writing $$
L_e =  \cP e^{i \int_e A \cdot dx} \sim e^{i A_\mu l} \qquad (l \to 0)
$$
where $\mu$ is in the direction of $e$ and $A_\mu$ is the continuous gauge field evaluated at $s(e)$ as before. 

In the case of the spectral action we obtain the following continuum limit.

\begin{prop}\label{prop:Scontlim}
Let $(A_v,H_v) = (M_N(\C), \C^N)$ for all vertices $v$. In the limit where $l \to  0$ (while $\Lambda \propto l^{-1}$), the spectral action 
$S_\Lambda$ becomes the action functional
\begin{align*}
\frac14 \int_M  \tr F_{\mu\nu}F^{\mu\nu} + 2 \int_M \tr (\partial_\mu \Phi - [iA_\mu,\Phi])(\partial^\mu \Phi - [iA^\mu,\Phi]) + 8\Lambda^2 \int_M \tr \Phi^2 +  \int_M \tr \Phi^4 .
\end{align*}
\end{prop}

\smallskip

\proof
For a plaquette as in Figure \ref{fig:plaquette} we find that 
$$
\tr \left( L_{\bar e_4} L_{\bar e_3} L_{e_2} L_{e_1} \right)
= \tr   e^{-il A_\nu(x)}  e^{-i l A_\mu(x+ l \hat \nu)}e^{i l A_\nu(x+ l \hat \mu)}e^{il A_\mu(x)} \sim \tr e^{il^2 F_{\mu\nu}}  \qquad (l \to 0)
$$
and similarly for $\tr \left( L_{\bar e_1} L_{\bar e_2} L_{e_3} L_{e_4} \right)$. 
As a consequence, in the limit $l \to 0$ (or $\Lambda \to \infty$) we have (modulo constant term)
$$
S_\Lambda \sim  \frac14 \int_M \tr F_{\mu\nu}F^{\mu\nu} 
$$
which is in concordance with the continuous derivation of the Yang--Mills action from the spectral action on a noncommutative manifold \cite{CC96}.

Concerning the Higgs-field, we determine the continuum limit of the remaining terms in $S_\Lambda[\{ L_e\}, \{D_v\}]$. With the vertex $v$ at position $x$ as in Figure \ref{fig:higgs-edge} we first note that
\begin{multline*}
\tr e^{-i A_\mu l} \Phi(x+l \hat \mu) e^{i A_\mu l} \Phi(x) \sim \tr \bigg( \Phi(x) \Phi(x+l \hat \mu) + l \Phi(x+l \hat \mu) [ i A_\mu, \Phi(x)] \\ - \frac12 l^2 [iA_\mu,\Phi(x+l \hat \mu)] [i A_\mu,\Phi(x)] \bigg) + \cO(l^3)
\end{multline*}
where $\Phi(x)$ is the continuous (hermitian) Higgs field corresponding to $D_x$ and $L_e$ is expanded in terms of $A_\mu$ as above. Substituting this in $S_\Lambda$, we find
\begin{align*}
S_\Lambda &= - \frac14 \sum_{\partial p =\bar e_4  \bar e_3 e_2 e_1} \left(\tr \left( L_{\bar e_4} L_{\bar e_3} L_{e_2} L_{e_1} \right)
+ \tr \left( L_{\bar e_1} L_{\bar e_2} L_{e_3} L_{e_4} \right)\right)\\
&\quad + \sum_v  l^4 \tr D_v^4 + 4 l^2 \sum_e \left( \tr D_{s(e)}^2 + \tr D_{t(e)}^2 - \tr L_e^* D_{t(e)} L_e D_{s(e)} \right)\\
& \sim \frac12 \tr e^{i l^2 F_{\mu\nu}} + l^4 \tr \Phi^4(x) + 2l^2 \sum_{\mu}  \tr \Phi^2(x) + \tr \Phi^2 (x+l \hat \mu) \\
& \quad + 2 l^4 \sum_\mu  \frac1{l^2} \tr(\Phi(x+l \hat \mu)- \Phi(x))^2\\& \quad - \frac2l \tr \Phi(x+l \hat \mu) [iA_\mu(x), \Phi(x)]  +  \tr  ([iA_\mu(x), \Phi(x)])^2
\end{align*}
modulo $\cO(l^3)$.  Thus, we obtain the continuum limit as in the statement.
\endproof

\begin{rem}
Note the above sign in the charge of the Higgs field, as compared to the usual convention; this is due to the fact that we consider the holonomies of our gauge fields $L_e$ to map from $s(e)$ to $t(e)$. In contrast, in \cite{DJK84} the unitaries $L_e$ on the edges transform as $L_e \to g_{s(e)} L_e g_{t(e)}^{-1}$ (compare with Proposition \ref{prop:dirac-gauge}). 
\end{rem}

Finally, the fermionic action as defined above by 
$$
S_F[\{ \psi_v\},\{ L_e\}, \{ D_v\}] = \left \langle \psi, D_{\Gamma,L} \psi \right \rangle
$$
coincides with the action for fermions on a lattice, coupled to a gauge field and a Higgs field. With the help of Equation \eqref{eq:dirac-limit}, it can be readily checked that $S_F$ gives rise to the usual fermionic action in the continuum limit. 

\begin{rem}
We might also define a topological action as in \cite{BoeS10,CC10} as
$$
S_{\textup{top}}[L] = \tr \gamma f(D_{\Gamma,L})
$$
using the pullback of the grading $\gamma$ on $M$ to $\Gamma$; still denoted by $\gamma$.

The problem here is that if $f(x)= x^4$ then the trace selects cycles in $\Gamma$ of length 4, whereas the trace of the corresponding Dirac gamma matrices vanishes in this case:
$$
\tr \gamma \gamma_{\tilde \mu} \gamma_{\tilde \nu} \gamma_{-\tilde \mu} \gamma_{-\tilde \nu}  = \tr \gamma \gamma_{\tilde \mu} \gamma_{\tilde \nu} \gamma_{\tilde \mu} \gamma_{\tilde \nu} = 0
$$
An alternative definition might be taken from \cite{Woi83,PS91}, but then the connection to the spectral action is not so easy to see.
\end{rem}

\subsubsection{Kogut--Susskind Hamiltonian}

Consider now a three-dimensional lattice, so $M= \R^3$, with all $A_v = M_n(\C), H_v = \C^n$ while $D_v=0$. It can be obtained for instance from the previous four-dimensional lattice through temporal gauge fixing. 
A similar computation as appeared in the proof of Theorem \ref{thm:sa} shows that
\begin{prop}
On a three-dimensional lattice, with $L_e \in U(n)$ and $f(x)=x^4$ we have
$$
\tr f( D_{\Gamma,L}) \propto \sum_{\partial p = e_4 \cdots e_1} 
 \left( \tr \left( L_{\bar e_4} L_{\bar e_3} L_{e_2} L_{e_1} \right)
+ \tr \left( L_{\bar e_1} L_{\bar e_2} L_{e_3} L_{e_4} \right)\right) 
+ \text{const.}
$$
with the sum over plaquettes (cf. Figure \ref{fig:plaquette}). 
\end{prop}
This is precisely the interaction term in the Kogut--Susskind Hamiltonian $\bH_{\textup{KS}}$, so that with the Hamiltonian of Eq. \eqref{eq:hamiltonian} 
$$
\bH_{\textup{KS}} = \bH + \tr f( D_{\Gamma,L}) 
$$ 
with  $\tr f( D_{\Gamma,L}) $ a bounded multiplication operator on $L^2(\X/\G)$ (cf. the discussion below Proposition \ref{prop:sa-gauge}).

\subsection{A proposal for gauge foams}
We propose a noncommutative generalization of spin foams as higher-dimensional analogues of spin networks. The construction of {\it gauge foams} is such that taking a ``slice'' of the spin foam at a given ``time'' will then produce a gauge network. With this in mind, it is intuitively clear how gauge foams encode the {\it dynamics} of quantum noncommutative spaces, while gauge networks give the {\it kinematics}.

A natural way to arrive at spin foams is by computing the partition function for lattice gauge fields \cite{OP01}, expressing probability amplitudes as sums over spin foams (on a fixed graph/lattice). Their computation is essentially based on the fact that the path integral depends only on the `plaquette product' of group elements assigned to the four edges of a plaquette. We already encountered this before in Theorem \ref{eq:sa}. 

For simplicity, we only propose a definition for closed gauge foams. The generalization to gauge foams between two gauge networks is straightforward, and can be done as in \cite{DeMaZa}. 
In the following, by a two-complex we mean a simplicial complex with two-dimensional faces, one-dimensional edges, and zero-dimensional vertices, endowed with the usual boundary operator $\partial$, which assigns to a face the formal sum of its boundary edges with positive or negative sign according to whether the induced orientation from the face agrees or not with the orientation of the corresponding edge.

\begin{defn}
A \emph{gauge foam} $\Psi$ is the set of data $(\Sigma, (\tilde \rho_f)_f, (\BB_e , \tilde \iota_e)_e, (A_v, H_v)_v)$ where
\begin{enumerate}
\item $\Sigma$ is an oriented two-complex.
\item $(A_v,\lambda_v,H_v)$ is an object in the category $\cC_0^s$ for each vertex $v \in \Sigma^{(0)}$.
\item For each face $f$, $\tilde \rho_f$ is a representation of the unitary group $G_f$ generated by the product of unitary maps $L_e$ on the edges $e$ bounding the face $f$:
$$
G_f := \left\{ g_1 \cdots g_k :   g_i \text{ or }  g_i^{-1} \in \left( \U(\tilde A_{t(e_k)})L_{\BB_{e_k}} \right)^{\pm 1}  \cdots \left( \U(\tilde A_{t(e_1)})L_{\BB_{e_1}} \right)^{\pm 1} \right\},
$$
where the sign is $+1$ if the orientation of $e$ in $\partial f$ agrees with that of $f$, and $-1$ if not. 
\item For each edge $e$ in $\Sigma$, an intertwiner of $\U(\tilde A_{t(e)})$-representations:
$$
\tilde \iota_e: \bigotimes_{f': e \in \partial f'} \tilde \rho_{f'}|_{\U(\tilde A_{t(e)})} \to \bigotimes_{f: \bar e \in \partial{f}} \tilde \rho_{f}|_{\U(\tilde A_{t(e)})}.
$$
\end{enumerate}
\end{defn}

The fact that we allow for arbitrary algebras and Hilbert spaces at the vertices introduces an infinite-dimensional degeneracy. Nevertheless, as in \cite{OP01} one can derive that the path integral with Boltzmann factor given by the spectral action is a sum over gauge foams:
$$
Z_\Sigma = \sum_{ \begin{smallmatrix} (A_v, H_v)_v \\ (\BB_e,\rho_e)_e \\ (\tilde \rho_{f}) \end{smallmatrix} }
\prod_f \bA_f(\tilde \rho_f) \prod_e  \bA_e (\BB_e, \tilde \rho_{F(e)} , \tilde \iota_e) \prod_v \bA_v (\tilde \rho_{F(v)}, \tilde \iota_{E(v)} )
$$
where $F(e)$ are the faces adjacent to $e$, and $E(v)$ the edges adjacent to $v$. We postpone a detailed analysis of gauge foams for future work.

\end{document}